\documentclass[longauth]{aa} 

\usepackage{xcolor}  %
\usepackage{natbib}  %
\bibpunct{(}{)}{;}{a}{}{,} 

\usepackage[colorlinks = true,
            linkcolor = blue,
            urlcolor  = blue,
            citecolor = blue]{hyperref} %

\usepackage{graphicx}
\usepackage{txfonts}
\usepackage{tikz}
\usepackage{threeparttable}
\usepackage{graphicx}
\usepackage{txfonts}
\usepackage{lipsum}
\usepackage{tabularx}
\usepackage{booktabs}
\usepackage{subcaption}     
\usepackage{lscape}             
\usepackage{placeins}  
\usepackage{threeparttable}  
                                
\usepackage[
    starfontserif 
    ]{starfont}
\usepackage{amsmath}

\usepackage{xcolor} %

\usepackage[version=4]{mhchem} 
\usepackage{linenoaa}

\newcommand{\mgic}{$\mathrm{MGIC}_{\rm rv}$}
\newcommand{\kpar}{$\mathcal{K}_{\rm p}$}
\newcommand{\ksmooth}{$\mathcal{K}_{\rm s}$}
\newcommand{\lrv}{$\ln \mathcal{L}_\mathrm{rv}^\prime$}

\begin{document} 

   \title{Two extremely irradiated volatile-rich sub-Neptunes with companions in the TOI-426 and TOI-1839 systems}
   \subtitle{Insights into arrival and survival near the lower edge of the Neptunian desert}
   
   \titlerunning{Extremely irradiated volatile-rich sub-Neptunes in TOI-426 and TOI-1839}


\author{A.~Castro-Gonz\'{a}lez\inst{\ref{obs_geneva},\thanks{Corresponding author: \url{amadeo.castro-gonzalez@unige.ch}}}
    \and
    O. Barrag\'{a}n\inst{\ref{warwik_2},\ref{oxford}}
    \and
    D.~J.~Armstrong\inst{\ref{warwik_2},\ref{warwik}}
    \and
    A.~Aguichine\inst{\ref{UCSC},\ref{astron_mexico}}
    \and
    V. Bourrier\inst{\ref{obs_geneva}}
    \and
    D.~Ehrenreich\inst{\ref{obs_geneva}}
    \and
    E.~X.~Tao\inst{\ref{UCSC}}
    \and
    J.~Lillo-Box\inst{\ref{CAB_villafranca}}
    \and
    M.~R.~Standing\inst{\ref{ESAC}}
    \and 
    S.~G.~Sousa\inst{\ref{CAUP}, \ref{dep_astro_porto}}
    \and
    E.~Delgado-Mena\inst{\ref{CAUP}}
    \and
    A.~Moya\inst{\ref{dep_astro_valencia},\ref{obs_valencia}}
    \and
    V.~Adibekyan\inst{\ref{CAUP}, \ref{dep_astro_porto}}
    \and 
    M.~Lendl\inst{\ref{obs_geneva}}
    \and 
    C.~Hellier\inst{\ref{keele_uni}}
    \and
    S.~B.~Howell\inst{\ref{NASA_ames}}
    \and
    E.~Furlan\inst{\ref{caltech_ipac}}
    \and
    C.~Ziegler\inst{\ref{physics_austin}}
    \and
    A.~C.~M.~Correia\inst{\ref{cfisuc-coimbra},\ref{obs_paris}}
    \and 
    K.~Cui\inst{\ref{warwik_2},\ref{warwik}}
    \and 
    P.~Figueira\inst{\ref{IAA_CSIC}}
    \and 
    J.~M.~Jenkins\inst{\ref{NASA_ames}}
    \and
    M.~A.~Fetzner Keniger\inst{\ref{warwik_2},\ref{warwik}}
    \and
    B.~Mer\'{i}n\inst{\ref{ESAC}}
    \and 
    A.~Osborn\inst{\ref{warwik_2},\ref{warwik}}
    \and 
    L.~Parc\inst{\ref{obs_geneva}}
    \and
    F.~Pepe\inst{\ref{obs_geneva}}
    \and 
    P.~J.~Wheatley\inst{\ref{warwik_2},\ref{warwik}}
    \and
    J.~N.~Winn\inst{\ref{princeton_astro}}
    }

\institute{Observatoire Astronomique de l’Université de Genève, Chemin Pegasi 51b, CH-1290 Versoix, Switzerland\label{obs_geneva}
    \and
    Department of Physics, University of Warwick, Gibbet Hill Road, Coventry, CV4 7AL, United Kingdom\label{warwik_2}
    \and
    Sub-department of Astrophysics, Department of Physics, University of Oxford, Oxford OX1 3RH, United Kingdom\label{oxford}
    \and
    Centre for Exoplanets and Habitability, University of Warwick, Gibbet Hill Road, Coventry, CV4 7AL, United Kingdom\label{warwik}
    \and
    Department of Astronomy and Astrophysics, University of California, Santa Cruz, CA, USA\label{UCSC}
    \and
    Instituto de Astronomía, Universidad Nacional Autónoma de México, km 107 Carretera Tijuana-Ensenada, CP 22860 Ensenada, Baja California, México\label{astron_mexico}
    \and
    Centro de Astrobiolog\'{i}a, CSIC-INTA, ESAC campus, 28692 Villanueva de la Ca\~{n}ada, Madrid, Spain\label{CAB_villafranca} 
    \and
    European Space Agency (ESA), European Space Astronomy Centre (ESAC), Camino Bajo del Castillo s/n, 28692 Villanueva de la Cañada, Madrid, Spain\label{ESAC}
    \and
    Instituto de Astrof\'isica e Ci\^encias do Espa\c{c}o, Universidade do Porto, CAUP, Rua das Estrelas, 4150-762 Porto, Portugal\label{CAUP}
    \and
    Departamento de F\'isica e Astronomia, Faculdade de Ci\^encias, Universidade do Porto, Rua do Campo Alegre, 4169-007 Porto, Portugal\label{dep_astro_porto}
    \and
    Departamento de Astronom\'ia y Astrof\'isica, Universitat de Val\`encia, Av. Vicent Andr\'es Estell\'es, 19, 46120 Burjassot, Spain\label{dep_astro_valencia}
    \and
    Observatori Astron\`omic, Universitat de Val\`encia, C/ Cat. Jos\'e Beltr\'an 2, 46980 Paterna, Spain\label{obs_valencia}
    \and
    Astrophysics Group, Keele University, Keele ST5 5BG, United Kingdom\label{keele_uni}
    \and
    NASA Ames Research Center, Moffett Field, CA 94035, USA\label{NASA_ames}
    \and
    NASA Exoplanet Science Institute, Caltech/IPAC, Mail Code 100-22, 1200 E. California Blvd., Pasadena, CA 91125, USA\label{caltech_ipac}
    \and
    Department of Physics, Engineering and Astronomy, Stephen F. Austin State University, 1936 North St, Nacogdoches, TX 75962, USA\label{physics_austin}
    \and
    CFisUC, Departamento de F\'isica, Universidade de Coimbra, 3004-516 Coimbra, Portugal\label{cfisuc-coimbra}
    \and
    Laboratoire Temps Espace, Observatoire de Paris, Universit\'e PSL, Sorbonne Universit\'e, CNRS, 75014 Paris, France\label{obs_paris}
    \and
    Instituto de Astrof\'isica de Andaluc\'ia-CSIC, Glorieta de la Astronom\'ia s/n, 18008 Granada, Spain\label{IAA_CSIC}
    \and
    Department of Astrophysical Sciences, Princeton University, 4 Ivy Lane, Princeton, NJ 08544, USA\label{princeton_astro}
    }

\date{Received 25 May 2026 / Accepted 2 September 2026}

 
  \abstract
    {The origin of the lower edge of the Neptunian desert remains an open question. The leading hypotheses are tidal disruption of volatile-rich sub-Neptunes following high-eccentricity tidal migration (HEM) and atmospheric escape of their primordial \ce{H2}/He envelopes.}
    {We aimed to confirm and characterise two TESS candidates near the lower edge of the desert, TOI-426.01 ($P_{\rm orb}\approx1.32$~d) and TOI-1839.01 ($P_{\rm orb}\approx1.42$~d), orbiting the solar-type stars TOI-426 (HD~34390; $V$ = 10.1 mag, G2~V) and TOI-1839 (TYC~304-865-1; $V$ = 11.0 mag, G9~V).}
    {We acquired 77 HARPS radial velocities (RVs) of TOI-426 and 70 of TOI-1839, and we performed a joint Bayesian analysis of the RVs and TESS photometry. The RVs of both targets show significant activity signals, which we modelled using shared-timescale and multidimensional Gaussian process frameworks to disentangle stellar variability and derive precise planetary parameters.}
    {We confirm TOI-1839~b ($R_{\rm p}=2.27 \pm 0.10\,\rm R_{\oplus}$, $M_{\rm p}=7.10 \pm 0.78\,\rm M_{\oplus}$) and TOI-426~b ($R_{\rm p}=2.19 \pm 0.09\,\rm R_{\oplus}$, $M_{\rm p}=6.7 \pm 1.8\,\rm M_{\oplus}$), and we detect an additional transiting sub-Neptune, TOI-1839~c ($P_{\rm orb}\approx4.02$ d, $R_{\rm p}=2.50 \pm 0.13\,\rm R_{\oplus}$, $M_{\rm p}=6.0 \pm 1.1\,\rm M_{\oplus}$), and a long-period giant planet, TOI-426~c (assuming a circular orbit, $P_{\rm orb}=235.8^{+9.2}_{-8.5}$ d, $M_{\rm p}\sin i=444^{+18}_{-17}\,\rm M_{\oplus}$). TOI-426~b ($F \approx 1700\,\rm F_{\oplus}$) and TOI-1839~b ($F \approx 1050\,\rm F_{\oplus}$) are among the most highly irradiated sub-Neptunes known to date, but their bulk densities nevertheless require a significant volatile component. For the three transiting planets, our interior-structure and atmospheric-escape models yield short mass-loss timescales of $\sim1$--100 Myr for \ce{H2}/He-dominated atmospheres but much longer timescales of $\sim100$--1000 Gyr for \ce{H2O}-dominated envelopes, favouring a steam-world scenario for these planets. We also explored the population of short-period low-mass sub-Neptunes with outer long-period giant companions and found a strikingly high incidence ($\sim35\%$) at the shortest orbital periods near the desert edge, where successful HEM outcomes are expected to circularise.}
    {The survival of highly irradiated volatile-rich sub-Neptunes near the lower edge of the Neptunian desert is difficult to reconcile with a classical picture in which this boundary is set solely by the evaporation of primordial \ce{H2}/He envelopes. At the same time, the high incidence of long-period giant companions is consistent with an important role for HEM in shaping this region.}

   \maketitle
%
\section{Introduction}

Over the past $\sim$20 years, exoplanet surveys have revealed a pronounced deficit of Neptunian planets at very short orbital periods, commonly referred to as the Neptunian desert \citep[][]{2005MNRAS.356..955M,2007A&A...461.1185L,2009MNRAS.396.1012D,2011ApJ...727L..44S,2011ApJ...742...38Y}. While this feature is most pronounced in the Neptunian domain ($R_{\rm p} \approx 4$--$8~R_{\oplus}$), it extends from the sub-Neptune ($R_{\rm p} \approx 1.8~R_{\oplus}$) to the Jovian ($R_{\rm p} \approx 20~R_{\oplus}$) regimes, where formation, migration, and atmospheric evolution theories predict fundamentally different outcomes \citep[e.g.][]{Dawson2018,2021JGRE..12606639B}. Today, it remains an open question whether the desert is produced by a single global mechanism or arises from the interplay of multiple processes acting across these regimes \citep[][]{2016A&A...589A..75M,Matsakos2016,2018Natur.553..477B,Bourrier2023,Bourrier2025,2018MNRAS.479.5012O,2022AJ....164..234V,2024A&A...689A.250C,2024A&A...691A.233C,2026A&A...709L..17C}.

A first step towards understanding the origin of the desert is to characterise its shape. \citet{2016A&A...589A..75M} identified an upper boundary with a negative slope in the giant-planet population in the period--radius plane (a related trend was previously reported in the period--mass plane by \citealt{2005MNRAS.356..955M}) and a lower boundary with a positive slope in the sub-Neptune population. Their extrapolated intersection at $P_{\rm orb} \approx 10$~d in the Neptunian domain defines the classical triangular desert shape. \citet{2024A&A...689A.250C} re-analysed the \textit{Kepler} sample, dominated by FGK hosts, correcting for observational biases. They found that the upper and lower boundaries are smoothly connected at $P_{\rm orb} \approx 3$~d in the Neptunian regime, forming a vertical boundary where the slope changes sign. This continuous geometry suggests that the different regimes may be physically connected, either through a single mechanism or through the interplay of closely coupled processes.

\citet{Matsakos2016} proposed the first unified view of the desert. They suggest that its upper and lower boundaries can be reproduced by high-eccentricity tidal migration \citep[HEM; e.g.][]{FordRasio2008,2011CeMDA.111..105C} of planets whose orbits were circularised near the tidal disruption limit \citep[e.g.][]{Roche1849,Ford2006}. Recently, \citet{2026A&A...709L..17C} revisited this scenario using updated mass--radius relations, finding good agreement with the revised desert geometry. Alternatively, atmospheric escape has also been invoked as a contributor \citep[e.g.][]{2004A&A...418L...1L}. In the sub-Neptune regime, atmospheric escape is expected to act efficiently \citep[e.g.][]{2012MNRAS.425.2931O,2013ApJ...776....2L} and has been shown to broadly reproduce the observed lower boundary of the desert \citep{2018MNRAS.479.5012O}. In this context, the strong evaporation of Neptunes that migrated early through the protoplanetary disk has been proposed as an additional pathway contributing to the desert shape \citep{Bourrier2023,Bourrier2025}. In contrast, the upper boundary has been shown to be stable against photoevaporation \citep[e.g.][]{2018MNRAS.476.5639I,2022AJ....164..234V}. Therefore, whether the desert is predominantly shaped by HEM or instead arises from a combination of HEM setting the upper boundary and atmospheric escape shaping the lower and Neptunian boundaries remains an open question. Other phenomena, such as post-circularisation orbital decay, Roche-lobe overflow, and tidal inflation, have also been discussed as potentially relevant in the desert region \citep[e.g.][]{2010ApJ...714L.222F,Matsakos2016,HallattMillholland2026,Hallatt2026,2026ApJ..1002L..30L}.

Additional insight into the origin of the desert can be obtained from the unusual planets found within it \citep[e.g.][]{2019MNRAS.486.5094W,2020NatAs...4.1148J,2025PASJ...77.1101I,2026A&A...707A...4C} and the population near its boundaries \citep[e.g.][]{2025AJ....169..147B,2025A&A...695A.281B,2026MNRAS.549f1958C,2025ApJ...989L..20J,2025MNRAS.541..919L,2025A&A...701A..79N,2025A&A...702A.209Y}. Some Neptune-sized planets in the desert exhibit unusually high densities ($\sim$5--10~$\mathrm{g\,cm^{-3}}$; \citealt{2020Natur.583...39A,2023Natur.622..255N}) consistent with being the remnant cores of tidally disrupted giants after HEM, although alternative scenarios have also been discussed \citep[e.g.][]{2020Natur.583...39A}. In the Jovian regime, the desert boundary coincides with the short-period edge of the hot-Jupiter pile-up, an overdensity of planets at periods of $\sim$3--5~d \citep[e.g.][]{2007ARA&A..45..397U}, thought to reflect the accumulation of successful HEM outcomes \citep[e.g.][]{Dawson2018}. A similar overdensity has been observed in the Neptunian domain, known as the Neptunian ridge, identified in \textit{Kepler} data \citep{2024A&A...689A.250C} and tentatively recovered in TESS samples \citep{2026MNRAS.546ag022C}, which has also been discussed as being linked to HEM \citep[e.g.][]{2018Natur.553..477B,Bourrier2023,Bourrier2025,2024A&A...691A.233C,Doyle2025,Yee2025,2026arXiv260709656A,2026ApJ...998..324H,2026arXiv260709451T,2026arXiv260620789Z}. In this context, \citet{2026A&A...709L..17C} showed that the tidal survival band, where HEM outcomes are expected to be preferentially deposited \citep[e.g.][]{FabryckyTremaine2007}, naturally traces both the pile-up and ridge extensions, providing a common explanation for the desert--overdensity structures in the Neptunian and Jovian domains. In contrast, no comparable overdensity has been observed in the sub-Neptune regime. There, the survival band is narrower and embedded within the high-occurrence small-planet population, making any excess more difficult to identify. Additional observables, such as the prevalence of companions capable of triggering HEM, may therefore provide further support for its role in shaping the lower desert boundary.

In this work, we investigate the lower edge of the Neptunian desert by confirming and characterising two TESS candidates in this region. We observed TOI-426 (HD~34390; $V$ = 10.1 mag, G2~V) and TOI-1839 (TYC~304-865-1; $V$ = 11.0 mag, G9~V) within the HARPS-NOMADS and NCORES collaborations (PI Armstrong), which use HARPS to perform radial velocity (RV) follow-up of candidates near the Neptunian desert detected by TESS \citep[e.g.][]{2023A&A...669A.109L,2023MNRAS.524.5804A,2025MNRAS.537.3175A,2023MNRAS.524.3877H,2024MNRAS.532.1612H}. These observations allowed us to confirm the desert-edge planets TOI-426~b and TOI-1839~b, measure their masses, and detect two additional companions, TOI-426~c and TOI-1839~c, providing insight into the orbital architecture of both systems. Section~\ref{sec:observations} describes the observations. Sections~\ref{sec:stellar_charact} and~\ref{sec:analysis_results} present the stellar characterisation and the transit and RV analysis. We discuss the results in Sect.~\ref{sec:discussion} and conclude in Sect.~\ref{sec:conclusions}.

\section{Observations}
\label{sec:observations}
\subsection{Transiting Exoplanet Survey Satellite}
\label{obs:tess}

\begin{figure}
    \centering
    \includegraphics[width=0.47\textwidth]{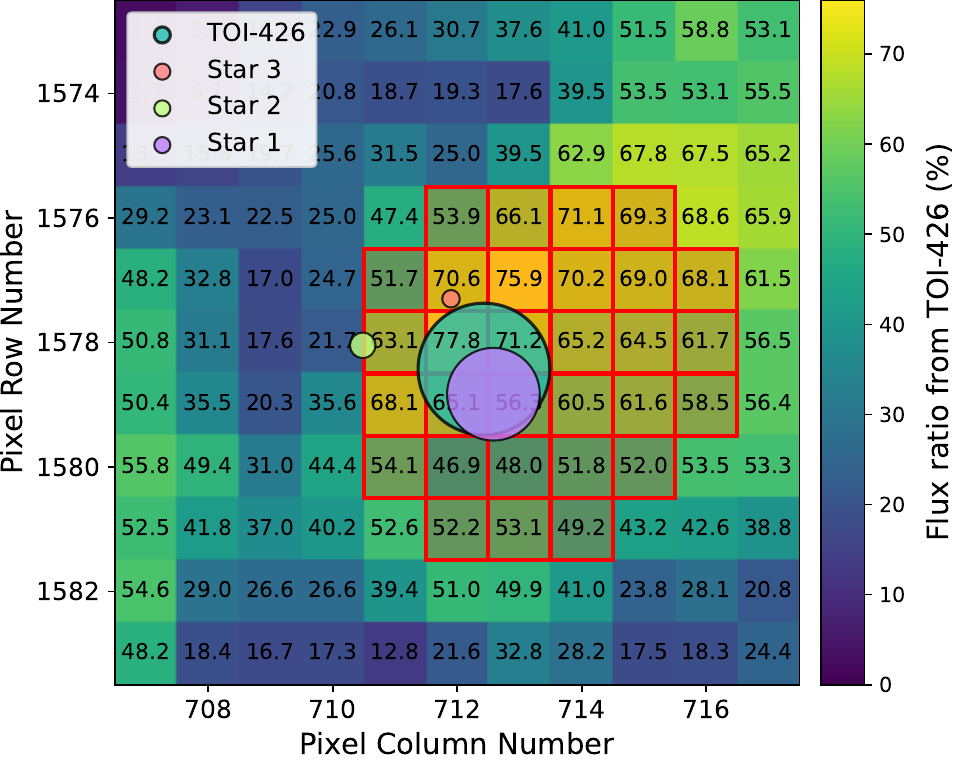}
    \caption{Heatmap showing the pixel-by-pixel flux fraction from TOI-426 in S5. The red grid indicates the photometric aperture. The disks represent the target star and the three main contaminant sources, with disk areas scaling with their fluxes. This plot was created with \href{https://github.com/castro-gzlz/TESS-cont}{\texttt{TESS-cont}} \citep{2024A&A...691A.233C}.}
    \label{fig:tess_cont_426}
\end{figure}

TOI-426 and TOI-1839 were observed by the Transiting Exoplanet Survey Satellite \citep[TESS;][]{2015JATIS...1a4003R} at a 2-min cadence. TOI-426 observations were carried out in sectors 5, 32, and 98 (hereinafter S5, S32, and S98), and TOI-1839 observations were carried out in sectors 23, 46, and 50 (hereinafter S23, S46, and S50), resulting in a total of 75\,767 target pixel files (TPFs) for TOI-426 and 47\,185 TPFs for TOI-1839. These TPFs are publicly available at the Mikulski Archive for Space Telescopes.\footnote{\url{https://mast.stsci.edu/portal/Mashup/Clients/Mast/Portal.html}} They were processed by the Science Processing Operations Center (SPOC) pipeline at the NASA Ames Research Center \citep{jenkinsSPOC2016}, which generated simple aperture photometry (SAP) and presearch data-conditioned simple aperture photometry \citep[PDCSAP;][]{2012PASP..124.1000S,2012PASP..124..985S}.

The SPOC photometric apertures were selected with the Create Optimal Apertures module \citep[COA;][]{2010SPIE.7740E..1DB,2020ksci.rept....7S}, which also provides the FLFRCSAP and CROWDSAP metrics used to correct the PDCSAP light curves for crowding. The CROWDSAP values indicate negligible contamination for TOI-1839 ($<0.1\%$), but almost 40\% contamination for TOI-426. We used \texttt{TESS-cont} \citep{2024A&A...691A.233C} based on \textit{Gaia} sources \citep{2018A&A...616A...1G,2023A&A...674A...1G} to identify the main contaminants. As shown in Fig.~\ref{fig:tess_cont_426}, the TOI-426 aperture is mainly contaminated by TIC 189013222 (Star 1; $G=10.8$ mag), which lies $8.9\arcsec$ from TOI-426 and contributes 89\% of the contaminant flux, with the remaining 11\% coming from TIC 189013227 (Star 2; $G=13.6$ mag), TIC 189013230 (Star 3; $G=14.4$ mag), and other sources.

In February 2019, the Transiting Planet Search \citep[TPS;][]{2002ApJ...575..493J,2010SPIE.7740E..0DJ,2020ksci.rept....9J} identified periodic flux decreases in the TOI-426 photometry with a periodicity of 1.3 d and a transit depth of $\approx 470$ ppm (parts per million). This signal was subjected to different diagnostic tests \citep{2018PASP..130f4502T,2019PASP..131b4506L}, including the SPOC difference-image centroiding test, which located TOI-426 within $-0.78 \pm 3.0$ arcsec of the transit source for the Sector 5--32 search. The TESS Science Office designated it as a TESS Object of Interest \citep[TOI-426.01;][]{2021ApJS..254...39G}. We used \texttt{TESS-cont} to determine whether the observed transit signal could have originated from any of the contaminant sources following \citet{2020MNRAS.499.5416C} and \citet{2021MNRAS.508..195D}. We obtained dilution-corrected transit depths of 956 ppm in Star 1, 2.3\% in Star 2, and 2.8\% in Star 3, so based on TESS data alone, we cannot discard these stars as possible sources of the observed signal. Notably, Star 1 is a TESS Object of Interest (TOI-949), with a 1.3-d candidate (TOI-949.01) that matches the ephemeris of TOI-426.01, indicating that both TOI-426.01 and TOI-949.01 were designated based on the same transit signal. We note that the high-resolution spectroscopy and imaging presented in Sects.~\ref{sec:obs_harps} and~\ref{sec:high_res_imaging} allowed us to confirm TOI-426 as the source of the signal and to rule out close stellar companions.

In May 2020, the SPOC pipeline identified a 1.4-d transit-like signal in the TOI-1839 photometry with a transit depth of $\approx 720$ ppm. The signal passed all diagnostic tests, including the SPOC difference-image centroiding test, which located TOI-1839 within $2.9 \pm 2.7$ arcsec of the transit source for the Sector 23--50 search, leading the TESS Science Office to designate it as TOI-1839.01. We also used \texttt{TESS-cont} to assess whether any contaminant source could be responsible for the flux decreases. We found dilution-corrected transit depths of 146\%, 452\%, and 455\% for the three main contaminant sources, indicating that TOI-1839.01 cannot originate from any nearby \textit{Gaia} source.

We analysed the SAP photometry after applying the SPOC crowding corrections. For TOI-426, we verified with \texttt{TESS-cont} that the aperture contamination from both \textit{Gaia} DR2 and DR3 is consistent with the SPOC dilution factors. We also tested the \texttt{default}, \texttt{hard}, and \texttt{hardest} \texttt{Lightkurve} quality masks \citep{2018ascl.soft12013L}, as well as the PDCSAP photometry, obtaining consistent results. We therefore adopted the SAP photometry with the \texttt{default} mask for the fiducial analysis. Tables~\ref{tab:TESS_data_426} and~\ref{tab:TESS_data_1839} list the TESS photometric data.

\subsection{High Accuracy Radial velocity Planet Searcher}
\label{sec:obs_harps}

We observed TOI-426 and TOI-1839 with the High Accuracy Radial velocity Planet Searcher \citep[HARPS;][]{2003Msngr.114...20M}. HARPS is a fibre-fed, cross-dispersed echelle spectrograph mounted on the ESO 3.6-m telescope at La Silla Observatory, Chile. It provides a wavelength coverage of 378--691\,nm and a resolving power of $R \approx 115{\,}000$, enabling RV measurements with precisions at the $\mathrm{m\,s^{-1}}$ level.

We acquired 78 spectra of TOI-426 between 25 September 2022 and 30 March 2023, and 71 spectra of TOI-1839 between 9 January 2021 and 18 June 2023, yielding a total of 149 HARPS observations. We excluded one spectrum of TOI-426 (JD 2459949.7) and one of TOI-1839 (JD 2459272.9) owing to their anomalously low S/N per pixel of 8.9 and 17.1, respectively. The retained TOI-426 spectra have S/N per pixel ranging from 31.5 to 75.3, with a median of 54.0, while the retained TOI-1839 spectra range from 27.4 to 52.8, with a median of 40.2. The TOI-426 spectra were obtained as part of the HARPS-NOMADS programme 108.21YY.001 (PI: Armstrong), while the TOI-1839 data were acquired under the HARPS-NOMADS programmes 108.21YY.001 (30 observations) and 108.21YY.002 (7 observations), as well as under the NCORES large programme 1102.C-0249(F) (34 observations; PI: Armstrong). All observations were performed in High Accuracy Mode with exposure times of 1800 s. The primary fibre was centred on the star, and the secondary fibre was placed on the sky to monitor the background light.

We reduced the raw spectra through the HARPS Data Reduction Software (\texttt{DRS}). The \texttt{DRS} extracts the RVs based on a modified version of the cross-correlation technique \citep{1996A&AS..119..373B}, where different spectral lines of a numerical mask are weighed according to their RV information \citep{2002A&A...388..632P}. We also used the HARPS \texttt{DRS} to extract several activity indicators such as the full width at half maximum (FWHM) and contrast of the
cross-correlation function (CCF), the bisector span (BIS), the H$\alpha$, Ca II H\&K doublet (Ca), and sodium doublet (NaD) line depths, and the $S$ index. In Tables~\ref{tab:HARPS_data_426} and~\ref{tab:HARPS_data_1839}, we present the complete HARPS data set for TOI-426 and TOI-1839, respectively. The HARPS RV uncertainties of TOI-426 range from 1.4 to 3.2 $\rm m\,s^{-1}$ (median of 1.9 $\rm m\,s^{-1}$). These small values contrast with the large RV variations, which show a peak-to-peak amplitude of 156 $\rm m\,s^{-1}$ and a standard deviation of 34 $\rm m\,s^{-1}$. The RV uncertainties of TOI-1839 range from 1.5 to 3.0 $\rm m\,s^{-1}$ (median of 1.9 $\rm m\,s^{-1}$), which are also considerably smaller than the observed RV peak-to-peak amplitude of 49 $\rm m\,s^{-1}$ and standard deviation of 13 $\rm m\,s^{-1}$. These large differences between the RV uncertainties and dispersions suggest the presence of prominent planetary or stellar signals.

\subsection{High-resolution imaging}
\label{sec:high_res_imaging}

High-resolution imaging is a crucial part of planet confirmation, as close stellar companions can contaminate the observed transit and bias the inferred stellar and planetary properties \citep[][]{2015ApJ...805...16C,2017A&A...606A..75C,2020ApJ...898...47F,2021AJ....162...75L,2021FrASS...8...10H,2022MNRAS.509.1075C}. To assess the presence of nearby companions, high-resolution imaging of TOI-426 and TOI-1839 was obtained with Gemini and SOAR in the context of the ExoFOP follow-up programme. The raw data, processed contrast curves, and ancillary diagnostics are available at the ExoFOP-TESS repository.\footnote{\url{https://exofop.ipac.caltech.edu/tess/}}

\subsubsection{Gemini}

TOI-426 was observed on 12 October 2019 and TOI-1839 on 9 June 2020 with the ‘Alopeke speckle instrument on the Gemini North 8-m telescope \citep{2021FrASS...8..138S}. ‘Alopeke provides simultaneous speckle imaging in two bands (562\,nm and 832\,nm) with output data products including reconstructed images and robust magnitude contrast limits on companion detection. Thirteen sets of 1000 $\times$ 0.06 s exposures were collected for TOI-1839, and five similar sets were collected for TOI-426. All images were subjected to Fourier analysis using the standard reduction pipeline \citep{2011AJ....142...19H}. No close companions were detected around either TOI-426 or TOI-1839 within the diffraction limit (20 mas) out to $1.2\arcsec$, down to contrast limits of 5--8\,mag relative to the target stars. %

\subsubsection{SOAR}

We also searched for stellar companions using speckle imaging with the 4.1-m Southern Astrophysical Research (SOAR) telescope \citep{2018PASP..130c5002T}, observing TOI-426 on 17 March 2019 and TOI-1839 on 27 February 2021. Observations were performed in Cousins $I$ band, a similar bandpass to TESS \citep{2020AJ....159...19Z}. Both observations reached a $5\sigma$ sensitivity to companions 4.8 magnitudes fainter than the targets at an angular separation of $1\arcsec$. No stars were detected within $3\arcsec$.

\subsection{Wide Angle Search for Planets}
\label{sec:obs_wasp}

TOI-426 and TOI-1839 were extensively monitored by the Wide Angle Search for Planets survey \citep[WASP;][]{2006PASP..118.1407P,2011EPJWC..1101004H}, which operated arrays of wide-field cameras in both the Northern and Southern hemispheres. For most of the survey, each array comprised eight cameras equipped with 200-mm f/1.8 lenses, 2048$\times$2048 CCDs, and 400--700 nm bandpass filters. In 2012, the WASP-South array was changed to 85-mm f/1.2 lenses with SDSS-$r$ filters, enabling a focus on brighter targets ($V \sim 6$--9 mag) without reaching saturation. TOI-426 was observed from 2008 to 2014 (58\,569 data points), and TOI-1839 from 2009 to 2014 (35\,522 data points), with yearly campaigns of $\sim$160 and $\sim$140 d, respectively. We used these long-baseline photometric data to search for rotation-induced stellar activity signals, as described in App.~\ref{sec:activity_wasp}. The corresponding WASP data sets are listed in Tables~\ref{tab:WASP_data_426} and~\ref{tab:WASP_data_1839}.

\section{Stellar characterisation}
\label{sec:stellar_charact}

\begin{table}[]
\tiny
\renewcommand{\arraystretch}{1.15}
\setlength{\tabcolsep}{2.6pt}
\caption{Stellar properties of TOI-426 and TOI-1839.}
\label{tab:stellar_properties}
\begin{tabular}{lllc}
\hline \hline
Parameter & TOI-426 & TOI-1839 & Source \\ \hline

\multicolumn{4}{l}{Identifiers} \\ \hline
BD                                      & -15 991                & ...                    & (1)       \\
HD                                      & 34390                  & ...                    & (2)       \\
ADS                                     & 3860 A                 & ...                    & (3)       \\
TYC                                     & 5902-387-1             & 304-865-1              & (4)       \\
Gaia DR2                                & 2983316311375470976    & 3729117900352224768    & (5)       \\
TIC                                     & 189013224              & 381714186              & (6)       \\ \hline

\multicolumn{4}{l}{Astrometric and photometric properties} \\ \hline
RA                                      & 05:16:24.53            & 13:07:19.95            & (7)       \\
Dec                                     & -15:30:36.9            & 05:51:07.3             & (7)       \\
$\mu_{\alpha}\cos\delta$ ($\mathrm{mas\,yr^{-1}}$) & $5.621 \pm 0.017$ & $-12.849 \pm 0.024$ & (7) \\
$\mu_{\delta}$ ($\mathrm{mas\,yr^{-1}}$) & $-25.478 \pm 0.016$ & $-61.347 \pm 0.020$ & (7) \\
$\varpi$ (mas)                          & 8.863 $\pm$ 0.018      & 8.542 $\pm$ 0.018      & (7)       \\
$\gamma$ ($\mathrm{km\,s^{-1}}$)       & $-13.99 \pm 0.23$      & $-8.16 \pm 0.36$       & (7)       \\
$\log(L_{\star}/L_{\odot})$             & $-0.022^{+0.012}_{-0.013}$ & $-0.264^{+0.011}_{-0.015}$ & (5) \\
$B$ (mag)                               & $10.933 \pm 0.010$     & $11.737 \pm 0.050$     & (8)       \\
$V$ (mag)                               & $10.116 \pm 0.010$     & $11.013 \pm 0.110$     & (8)       \\
$T$ (mag)                               & $9.5969 \pm 0.0062$    & $10.2173 \pm 0.0061$   & (6)       \\ \hline

\multicolumn{4}{l}{Atmospheric parameters and spectral type} \\ \hline
$T_{\rm eff}$ (K)         & $5750 \pm 62$     & $5347 \pm 67$     & Sect.~\ref{subsec:atmospheric_parameters} \\
$\log g$ (dex)            & $4.43 \pm 0.11$   & $4.35 \pm 0.12$   & Sect.~\ref{subsec:atmospheric_parameters} \\
$\log g_{\rm Gaia}$ (dex) & $4.47 \pm 0.03$   & $4.51 \pm 0.03$   & Sect.~\ref{subsec:atmospheric_parameters} \\
$[\mathrm{Fe/H}]$ (dex)   & $0.16 \pm 0.04$   & $0.09 \pm 0.05$   & Sect.~\ref{subsec:atmospheric_parameters} \\
$\xi_{\mathrm{t}}$ ($\mathrm{km\,s^{-1}}$) & $1.08 \pm 0.03$ & $0.88 \pm 0.05$ & Sect.~\ref{subsec:atmospheric_parameters} \\
Sp. type$^{(a)}$               & G2 V              & G9 V              & Sect.~\ref{subsec:atmospheric_parameters} \\ \hline

\multicolumn{4}{l}{Physical parameters} \\ \hline
$R_{\star}$ ($\mathrm{R_{\odot}}$) & $0.99 \pm 0.03$ & $0.89 \pm 0.03$ & Sect.~\ref{subsec:atmospheric_parameters} \\
$M_{\star}$ ($\mathrm{M_{\odot}}$) & $1.00 \pm 0.02$ & $0.86 \pm 0.01$ & Sect.~\ref{subsec:atmospheric_parameters} \\
$v \sin i_{\star}$ ($\mathrm{km\,s^{-1}}$) & $3.5 \pm 0.1$ & $1.4 \pm 0.1$ & Sect.~\ref{subsec:atmospheric_parameters} \\
$\mathrm{Age}_{\rm gyro}$ (Gyr) & $1.57^{+1.70}_{-0.61}$ & $4.1^{+1.6}_{-1.4}$ & Sect.~\ref{subsec:age} \\
$\mathrm{Age}_{\rm CC}$ (Gyr)   & $0.94^{+0.75}_{-0.72}$ & $3.4^{+1.8}_{-1.9}$ & Sect.~\ref{subsec:age} \\
$P_{\rm rot}$ (d)               & $12.37^{+0.10}_{-0.09}$ & $24.00^{+0.47}_{-0.46}$ & Sect.~\ref{subsec:analysis_RV} \\ \hline

\multicolumn{4}{l}{Galactic space velocities and membership} \\ \hline
$U_{\rm LSR}^{(b)}$ ($\mathrm{km\,s^{-1}}$) & $-29.37 \pm 0.17$ & $-19.82 \pm 0.09$ & Sect.~\ref{subsec:atmospheric_parameters} \\
$V_{\rm LSR}^{(b)}$ ($\mathrm{km\,s^{-1}}$) & $3.45 \pm 0.13$   & $-22.08 \pm 0.12$ & Sect.~\ref{subsec:atmospheric_parameters} \\
$W_{\rm LSR}^{(b)}$ ($\mathrm{km\,s^{-1}}$) & $10.21 \pm 0.11$  & $-13.60 \pm 0.33$ & Sect.~\ref{subsec:atmospheric_parameters} \\
Gal. population & Thin disk (99\%) & Thin disk (99\%) & Sect.~\ref{subsec:atmospheric_parameters} \\
\hline
\end{tabular}

\tablefoot{\tiny References: (1) \citealp{1886BD....C......0S}; (2) \citealp{1918AnHar..91....1C}; (3) \citealp{1932ngcd.book.....A}; (4) \citealp{2000A&A...355L..27H}; (5) \citealp{2018A&A...616A...1G}; (6) \citealp{2019AJ....158..138S}; (7) \citealp{2023A&A...674A...1G}; (8) \citealp{2013AJ....145...44Z}. (a) The spectral types are inferred from the spectroscopic effective temperatures following \citet{2013ApJS..208....9P}. (b) The Galactic space velocities are referred to the local standard of rest (LSR) as estimated by \citet{2003A&A...409..523R}, and the probabilities are computed following \citet{2003A&A...410..527B}.}
\end{table}

TOI-426 and TOI-1839 are bright stars ($V = 10.1$ and 11.0\,mag, respectively; Table~\ref{tab:stellar_properties}) located in the solar neighbourhood at heliocentric distances of 113 and 117\,pc, respectively. TOI-426 has a co-moving companion at a separation of $9\arcsec$ (TIC 189013222; Star 1 in Fig.~\ref{fig:tess_cont_426}), previously reported in historical catalogues \citep[e.g.][]{1886BD....C......0S,1932ngcd.book.....A} and confirmed by \textit{Gaia} proper motions, parallaxes, and RVs. We found no evidence of additional co-moving companions around TOI-1839.

\subsection{Stellar atmospheric parameters and abundances}
\label{subsec:atmospheric_parameters}

The stellar atmospheric parameters ($T_{\mathrm{eff}}$, $\log g$, $\xi_{\mathrm{t}}$, $[\mathrm{Fe/H}]$) were derived using the \texttt{ARES+MOOG} methodology as described by \citet{Santos2013}, \citet{Sousa2014}, and \citet{Sousa2021}. For this analysis, we used the latest version of \texttt{ARES}\footnote{\url{https://github.com/sousasag/ARES}} \citep{Sousa2007,Sousa2015}. We combined the spectra of TOI-426 and TOI-1839 and measured the equivalent widths (EW) of the iron lines selected by \citet{Sousa2008}. The parameters were derived using ionisation and excitation equilibrium together with a grid of Kurucz model atmospheres \citep{Kurucz1993} and the radiative transfer code \texttt{MOOG} \citep{Sneden1973}. We also derived trigonometric surface gravities using \textit{Gaia} DR3 following \citet{Sousa2021}. The stellar masses and radii were inferred using the relations from \citet{Torres2010}. The derived $T_{\mathrm{eff}}$, $\log g$, $\xi_{\mathrm{t}}$, $[\mathrm{Fe/H}]$, $R_{\star}$, and $M_{\star}$ are listed in Table~\ref{tab:stellar_properties}. The photometry-based characterisation from the TESS Input Catalog \citep[TIC v8.2;][]{2019AJ....158..138S} is consistent within $1\sigma$ with our spectroscopic results.

We also determined the abundances of C, O, Mg, Si, Ti, Ni, and neutron-capture elements using the atmospheric parameters and the classical curve-of-growth method \citep{Adibekyan-15,Adibekyan-16,DelgadoMena2017,DelgadoMena2021}. Both stars exhibit abundance patterns typical of thin-disk stars in the solar neighbourhood \citep{Adibekyan-12}, consistent with their Galactic space velocities. In addition, we derived the lithium abundances through spectral synthesis following \citet{DelgadoMena2014}. The low lithium abundance of TOI-1839 is consistent with an old age, whereas the high lithium abundance of TOI-426 suggests a relatively young star \citep[e.g.][]{Rathsam2023}. All abundances are listed in Table~\ref{tab:chemical_abundances}.

\subsection{Age}
\label{subsec:age}

We estimated the stellar ages through gyrochronology using the empirical relations from \citet{2019JOSS....4.1469A,2019AJ....158..173A}, adopting the stellar atmospheric parameters, \textit{Gaia} parallaxes, and rotation periods listed in Table~\ref{tab:stellar_properties}. The rotation periods were inferred from the activity modelling of the HARPS RVs and activity indicators in Sect.~\ref{subsec:analysis_RV}. These periods are consistent with those obtained from a complementary frequency analysis of the WASP photometry, presented in App.~\ref{sec:activity_wasp}. We also estimated the ages using chemical clocks based on the [Y/Mg], [Sr/Mg], [Y/Si], [Y/Ti], and [Y/Zn] abundance ratios following \citet{2019A&A...624A..78D,2022A&A...660A..15M,2024MNRAS.528.3464R}. The resulting gyrochronology and chemical-clock age estimates are listed in Table~\ref{tab:stellar_properties}. Both methods consistently indicate that TOI-426 is younger than TOI-1839, also in agreement with the lithium abundances.

\section{Analysis and results}
\label{sec:analysis_results}

\begin{figure*}
    \centering
    \includegraphics[width=0.373157\textwidth]{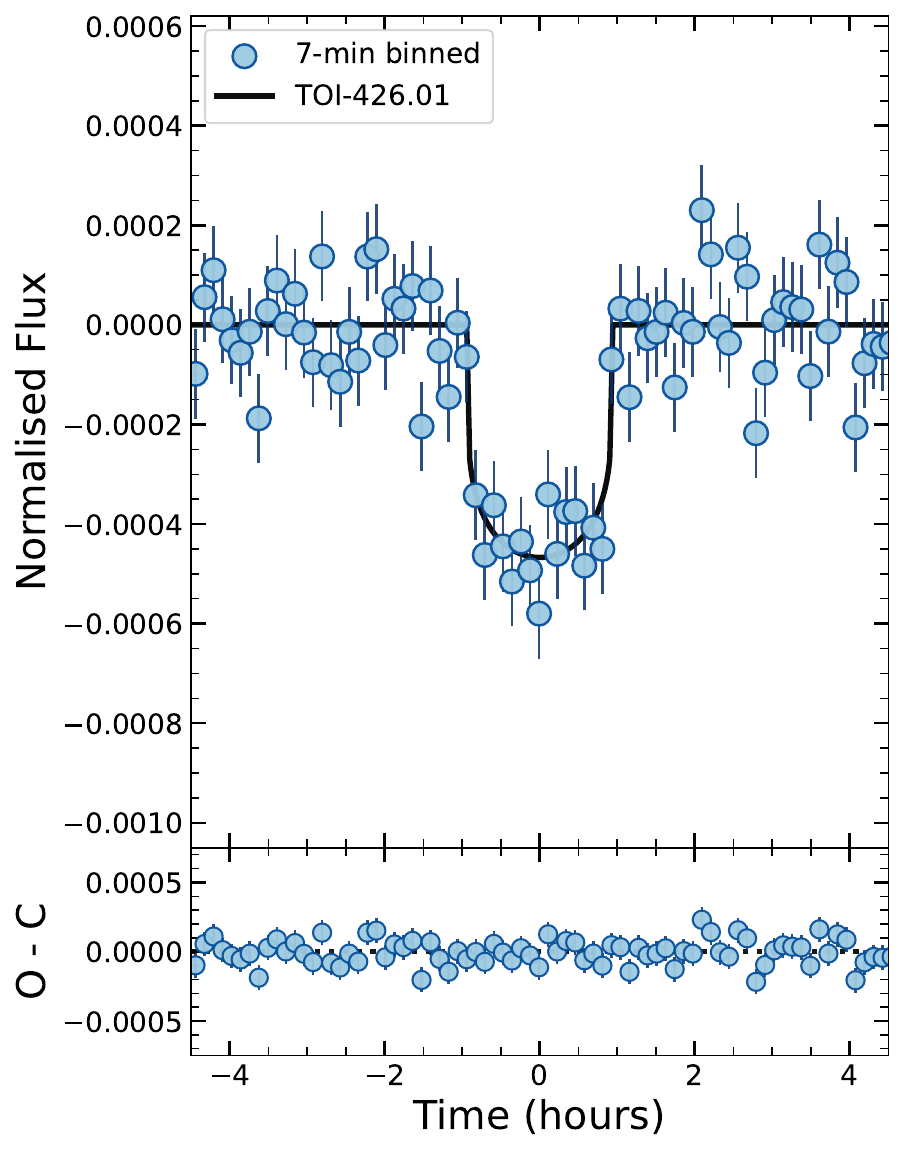}
    \includegraphics[width=0.3005\textwidth]{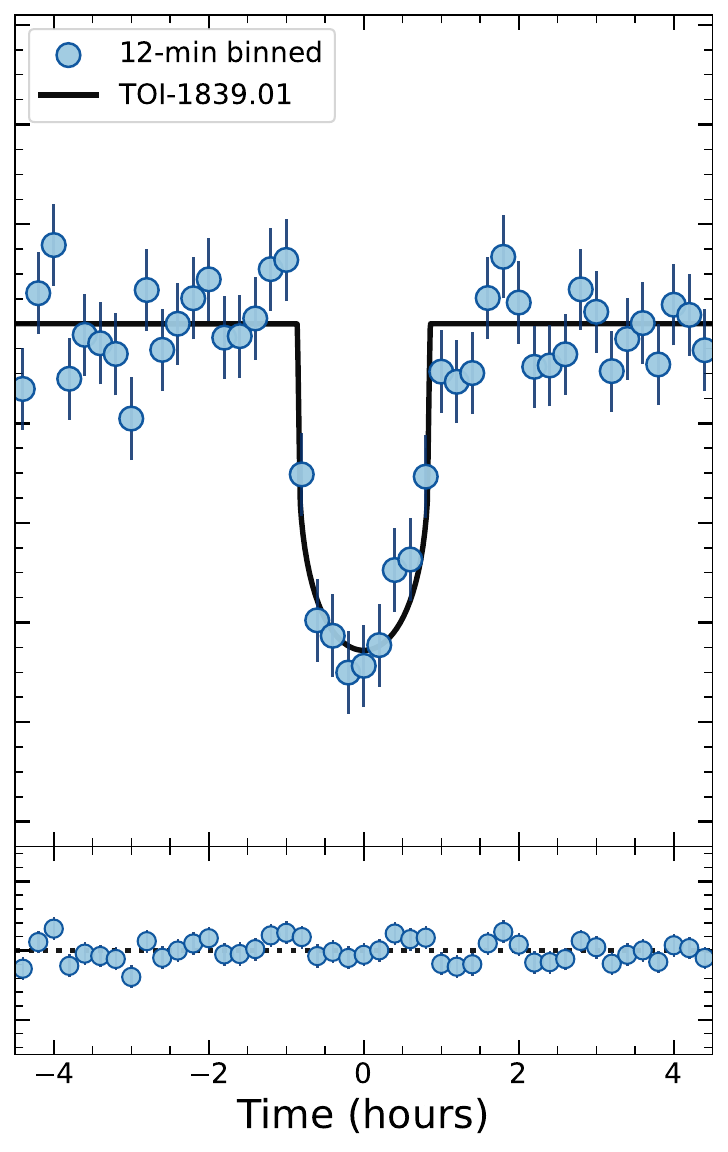}
    \includegraphics[width=0.3005\textwidth]{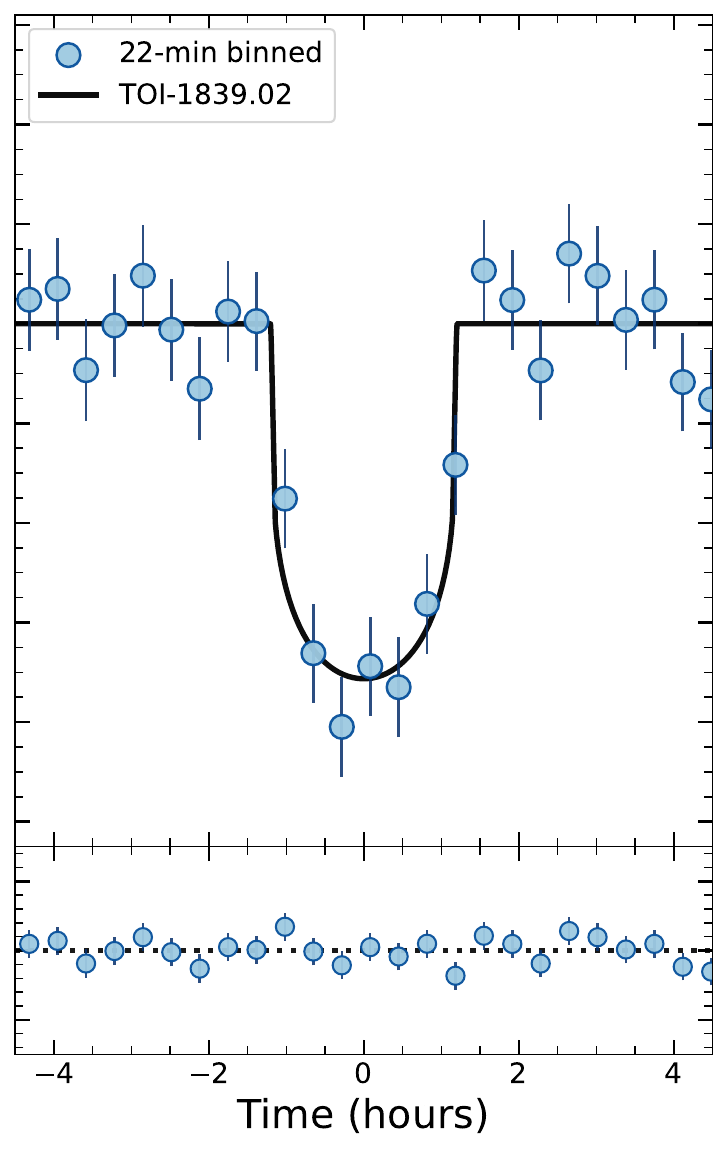}
    \caption{TESS SAP photometry of TOI-426 and TOI-1839 after subtraction of the correlated noise components and phase-folded at the inferred orbital periods: $1.32$~d for TOI-426.01, $1.42$~d for TOI-1839.01, and $4.02$~d for TOI-1839.02. The black solid lines represent the median posterior transit models.}
    \label{fig:transits_folded}
\end{figure*}

Before carrying out the joint transit and RV modelling, we performed a frequency analysis of the TESS photometry, WASP photometry, and HARPS RVs and activity indicators, which is presented in App.~\ref{sec:frequency}. This analysis independently recovers the TESS candidates, reveals the additional 4.02-d transiting signal in TOI-1839, and confirms the presence of rotation-related stellar-activity signals. Owing to the large volume of TESS and HARPS data, our modelling was then performed in two steps. In Sect.~\ref{subsec:analysis_TESS}, we de-trended the TESS photometry by jointly modelling the correlated noise and the transit signals to better preserve the transit shapes. In Sect.~\ref{subsec:analysis_RV}, we analysed the de-trended photometry together with the HARPS RVs. In Sect.~\ref{sec:sensitivity_limits}, we derive sensitivity limits from the RV residuals.

\subsection{TESS photometry modelling and de-trending}
\label{subsec:analysis_TESS}

To model the transit signals, we used the \texttt{batman} implementation \citep{2015PASP..127.1161K} of the quadratic limb-darkened model by \citet{2002ApJ...580L.171M}. The model parameters are the orbital period ($P_{\rm orb}$), time of inferior conjunction ($T_{0}$), orbital inclination ($i$), quadratic limb darkening (LD) coefficients ($u_{1}$, $u_{2}$), planet-to-star radius ratio ($R_{\rm p}/R_{\star}$), and scaled semi-major axis ($a/R_{\star}$). We re-parametrised the LD following \citet{2013MNRAS.435.2152K} for efficient sampling: $q_{1} = (u_{1} + u_{2})^{2}$ and $q_{2} = 0.5\,u_{1}(u_{1} + u_{2})^{-1}$. The scaled semi-major axis was parametrised in terms of the stellar density ($\rho_{\rm \star}$) via Kepler’s third law \citep[e.g.][]{2007ApJ...664.1190S,2010exop.book...55W}, and the orbital inclination was parametrised via the impact parameter \citep[$b$;][]{2003ApJ...585.1038S}.

To model the correlated noise, we included a Gaussian process (GP) regression component \citep{2006gpml.book.....R,2012RSPTA.37110550R} defined by the approximate Matérn-3/2 kernel as implemented in \texttt{celerite} \citep{2017AJ....154..220F}. Owing to its flexibility, this kernel has been widely used to model TESS photometry affected by a mixture of stellar variability and instrumental systematics \citep[e.g.][]{2021A&A...656A.124K,2022AJ....163..298M,2023A&A...679A..33D,2023A&A...673A..32M,2023A&A...677A.182M,2024A&A...688A.216H,2024MNRAS.535..531S}. This kernel is expressed as a function of the absolute temporal separation between two data points, $\tau = |t_{i} - t_{j}|$, as

\begin{equation}
    K_{3/2} = \eta_{\sigma}^{2} \left[ \left(1 + \frac{1}{\epsilon} \right) e^{-(1-\epsilon) \sqrt{3} \tau / \eta_{\rho}} + \left(1 - \frac{1}{\epsilon} \right) e^{-(1+\epsilon) \sqrt{3} \tau / \eta_{\rho}} \right],
\end{equation}

\noindent where the hyperparameters $\eta_{\sigma}$ and $\eta_{\rho}$ represent the characteristic amplitude and timescale of the correlated variations, respectively.  The parameter $\epsilon$ controls the approximation to the exact Matérn-3/2 kernel and was fixed to its default value of $10^{-2}$. As the amplitudes and timescales of the TESS systematics may vary between sectors, we fitted these parameters independently ($\eta_{\sigma_{i}}$ and $\eta_{\rho_{i}}$, where $i$ denotes the sector). For each sector, we also included a jitter term ($\sigma_{\textrm{TESS},i}$) to account for uncorrelated noise.

We derived the posterior distributions of the model parameters using a dynamic nested sampling algorithm \citep{2004AIPC..735..395S,10.1214/06-BA127,2014A&A...564A.125B,2019S&C....29..891H} as implemented in \texttt{dynesty} \citep{2020MNRAS.493.3132S,sergey_koposov_2024_12537467}. We employed 10\,000 live points and adopted a conservative stopping criterion of $\Delta\mathcal{Z} < 10^{-7}$, where $\Delta\mathcal{Z}$ denotes the change in Bayesian evidence between successive iterations. We adopted uniform priors for all parameters except $\rho_{\rm \star}$, $q_{1}$, and $q_{2}$. The stellar density was constrained using a Gaussian prior based on our stellar characterisation (Sect.~\ref{sec:stellar_charact}; Table~\ref{tab:stellar_properties}). Quadratic LD coefficients were estimated with \texttt{ldtk} \citep{2015MNRAS.453.3821P}, using synthetic spectra from \citet{2013A&A...553A...6H}, the stellar spectroscopic parameters, and the TESS transmission curve. We adopted Gaussian priors centred on the \texttt{ldtk} values, with uncertainties inflated to $5\sigma$. Broader LD priors, including uncertainties of 0.2 \citep{2022AJ....163..228P} and uniform priors on $q_{1}$ and $q_{2}$, yielded consistent $R_{\rm p}/R_{\star}$ values within $1\sigma$. Models with individual transit timing offsets showed no significant TTVs or fit improvement, so we adopted the no-TTV model.

In Fig.~\ref{fig:transits_folded}, we show the de-trended (i.e. GP-subtracted) TESS photometry phase-folded at the inferred orbital periods of TOI-426.01, TOI-1839.01, and TOI-1839.02. In Fig.~\ref{fig:TESS_example_sectors}, we show example TESS sectors together with the corresponding median posterior models. In Table~\ref{tab:TESS_GP_hyperparams}, we list the median and $1\sigma$ intervals of the posterior distributions of the GP hyperparameters.

\subsection{TESS and HARPS joint modelling}
\label{subsec:analysis_RV}

We modelled the TOI-426 and TOI-1839 systems by jointly fitting the HARPS RVs and the de-trended TESS photometry. The same transit model, parametrisation, and priors described in Sect.~\ref{subsec:analysis_TESS} were adopted. For the RVs, we tested several models combining Keplerian signals and GPs to account for activity-induced RV signals. To include the planetary signals in the RV time series, we parametrised them using \texttt{radvel} \citep{2018PASP..130d4504F} through the orbital period ($P_{\rm orb}$), time of inferior conjunction ($T_{0}$), RV semi-amplitude ($K$), orbital eccentricity ($e$), and argument of periastron ($\omega$). We also included an RV offset ($\gamma$) representing the systemic velocity, and an RV jitter term ($\sigma_{\rm HARPS}$). We adopted uniform priors for most model parameters and modified Jeffreys priors for the jitter terms, which remain weakly informative over several orders of magnitude while preventing an unphysical preference for large noise amplitudes, as commonly done in Bayesian RV analyses \citep[e.g.][]{2005ApJ...631.1198G,2011MNRAS.415.2523G,2021MNRAS.507.1847R,2022MNRAS.517.5050R}.

\subsubsection{Stellar activity modelling with GPs}
\label{sec:GPs}

\begin{figure*}
    \centering
    \includegraphics[width=0.96\textwidth]{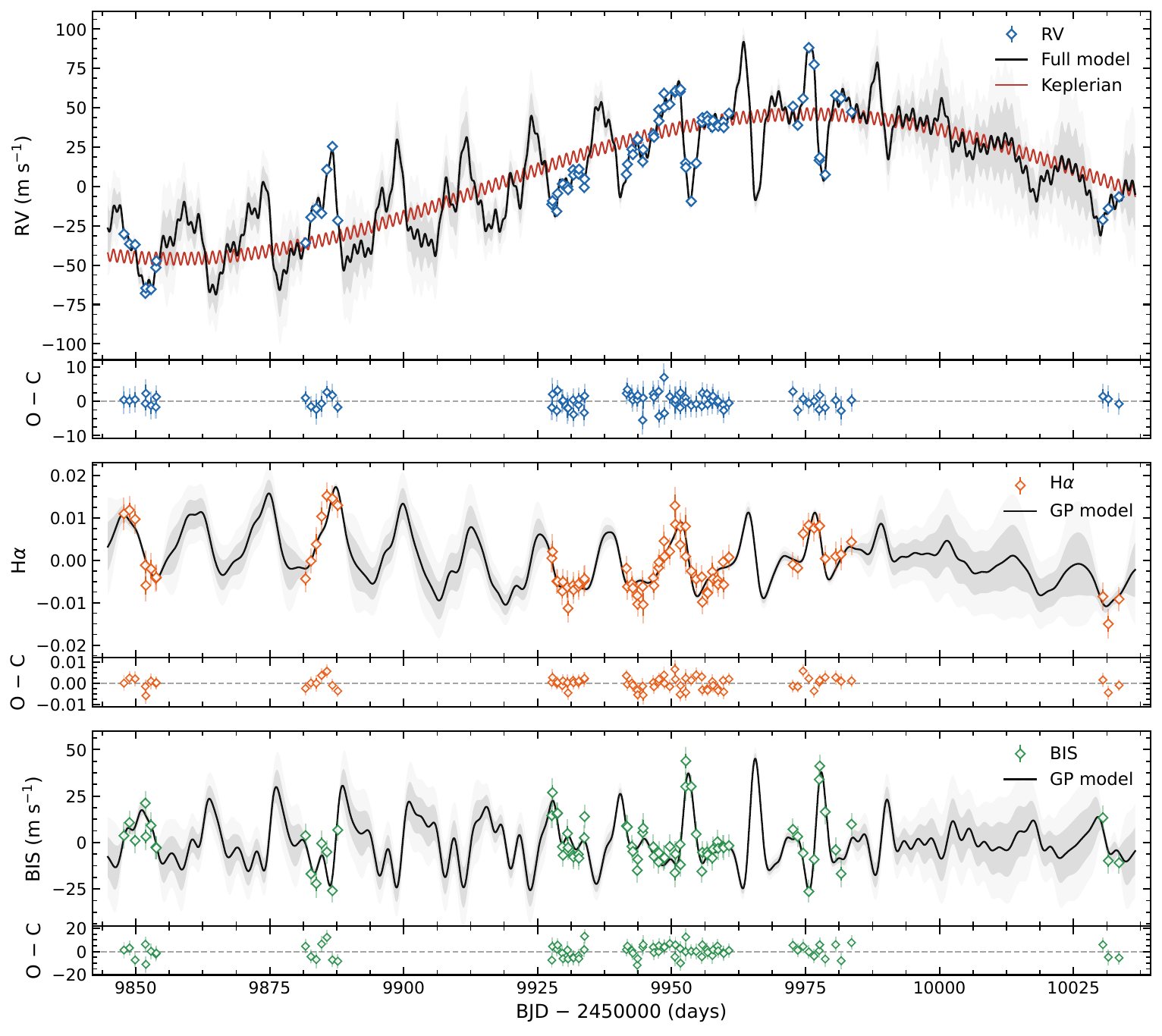}
    \caption{Full RV model for TOI-426. Top panel: HARPS RVs, with the black and red lines showing the median posterior full and Keplerian models, respectively. The Keplerian model corresponds to the sum of the TOI-426\,b and TOI-426\,c signals. Middle and bottom panels: HARPS H$\alpha$ and BIS, with the black line showing the median posterior GP model. In all panels, the time series were corrected for instrumental offsets, and the dark and light grey bands denote the $1\sigma$ and $2\sigma$ confidence intervals.}
    \label{fig:426_RV_full}
\end{figure*}

To model the stellar activity signals in the RV time series, we adopted two different GP regression frameworks. In both cases, we adopted a physically motivated quasi-periodic covariance function \citep{2015ITPAM..38..252A}, as activity signals are often neither strictly periodic nor fully coherent \citep[e.g.][]{2014ApJ...796..132D,2019ApJ...871..187N}. This kernel has been widely used to model activity-affected RV signals \citep[e.g.][]{2014MNRAS.443.2517H,2016A&A...588A..31F,figueira2025} and is expressed as
\begin{equation}
    K_{QP}(\tau) = A^2 \, \exp \left[
    - \frac{\tau^{2}}{2\lambda_{\rm e}^{2}}
    - \frac{\sin^{2}\left(\pi \tau P_{\rm GP}^{-1}\right)}{2\lambda_{\rm P}^{2}}
    \right],
    \label{eq:QPkernel}
\end{equation}

\noindent where $A$ sets the amplitude scale of the activity signal, $\lambda_{\rm e}$ is the decay timescale, $P_{\rm GP}$ is the characteristic periodic timescale (typically associated with the stellar rotation period, $P_{\rm GP} \equiv P_{\rm rot}$), and $\lambda_{\rm P}$ is the periodic coherence scale \citep{2014MNRAS.443.2517H,2016A&A...588A..31F,2018MNRAS.474.2094A}.

The first GP framework is a variation of the one-dimensional GP approach, where the RVs are modelled jointly with one activity indicator using shared timescale hyperparameters \citep[period and evolution timescales; e.g.][]{2020A&A...639A..77S,2020A&A...642A.121L,2022A&A...658A.115F}. We therefore refer to it as the shared-timescale GP (ST-GP) framework. This technique has been widely used to model activity-induced RV signals \citep[e.g.][]{2015ApJ...808..127G,2019ApJ...883...79D,2022A&A...665A.154B}, but it can produce overly flexible models in which planetary signals may be partially absorbed or distorted \citep[e.g.][]{2021MNRAS.503.1248A,2021MNRAS.507.1847R}. For the ST-GP analyses, we used the \texttt{dynesty} nested sampler to derive posterior distributions and Bayesian evidence, as in the photometric analysis (Sect.~\ref{subsec:analysis_TESS}).

The second framework we adopted is the multidimensional GP (multi-GP) framework by \citet{2015MNRAS.452.2269R}, which assumes that different time series are connected through a common latent function \citep[e.g.][]{2009A&A...495..959B,2012A&A...545A.109B,2012MNRAS.419.3147A,2014ApJ...796..132D}. In this approach, an $N$-dimensional GP model is constructed over $N$ time series $\mathcal{A}_i$, expressed as
\begin{equation}
\begin{aligned}
 \mathcal{A}_1 &= A_{1} G(t) + B_{1} \dot{G}(t), \\
 \vdots \\
 \mathcal{A}_N &= A_{N} G(t) + B_{N} \dot{G}(t),
\end{aligned}
\label{eq:gps}
\end{equation}
\noindent
where the coefficients $A_i$ and $B_i$ link each time series to the latent function $G(t)$ and its time derivative $\dot{G}(t)$. The process $G(t)$ is assumed to trace the temporal evolution of the projected surface coverage of stellar active regions \citep[e.g.][]{2015MNRAS.452.2269R,2022MNRAS.509..866B}. In this framework, activity indicators can be divided into RV-like indicators, which depend on both $G(t)$ and $\dot{G}(t)$, and photometry-like indicators, which depend only on $G(t)$ \citep{Barragan2023}. Throughout this work, we refer to configurations modelling the RVs alone, the RVs with one activity indicator, and the RVs with two activity indicators as 1D-GP, 2D-GP, and 3D-GP, respectively. The multi-GP approach has shown excellent performance for active stars \citep[e.g.][]{2019MNRAS.490..698B,2022MNRAS.514.1606B,2025A&A...702A..32H} and is therefore well suited to our analysis. We implemented this framework with \texttt{pyaneti} \citep{2019MNRAS.482.1017B,2022MNRAS.509..866B}, sampling the posterior parameter space with the ensemble Markov chain Monte Carlo algorithm based on the affine-invariant sampler of \citet{2013PASP..125..306F}. We used 250 walkers and retained the final 5000 iterations of the MCMC chains with a thinning factor of 10, resulting in 125\,000 posterior samples per parameter. Model comparison among the multi-GP configurations was performed using \mgic\ \citep{mgic}, an extension of the Akaike information criterion \citep[AIC;][]{Akaike1974} developed for RV analyses. Its likelihood term, \lrv, measures how well each model explains the RV data, while the penalty term combines the number of explicitly fitted parameters, \kpar, and the complexity induced by the GP component, \ksmooth. One advantage of \mgic\ over the AIC is that it enables consistent comparisons among multi-GP models with different dimensionalities and combinations of activity indicators. For completeness, we also repeated the model comparisons using the AIC and reached the same conclusions.

\subsubsection{Results on TOI-426}
\label{sec:TOI-426}

\begin{figure*}
    \centering

    \includegraphics[width=0.2723\textwidth]{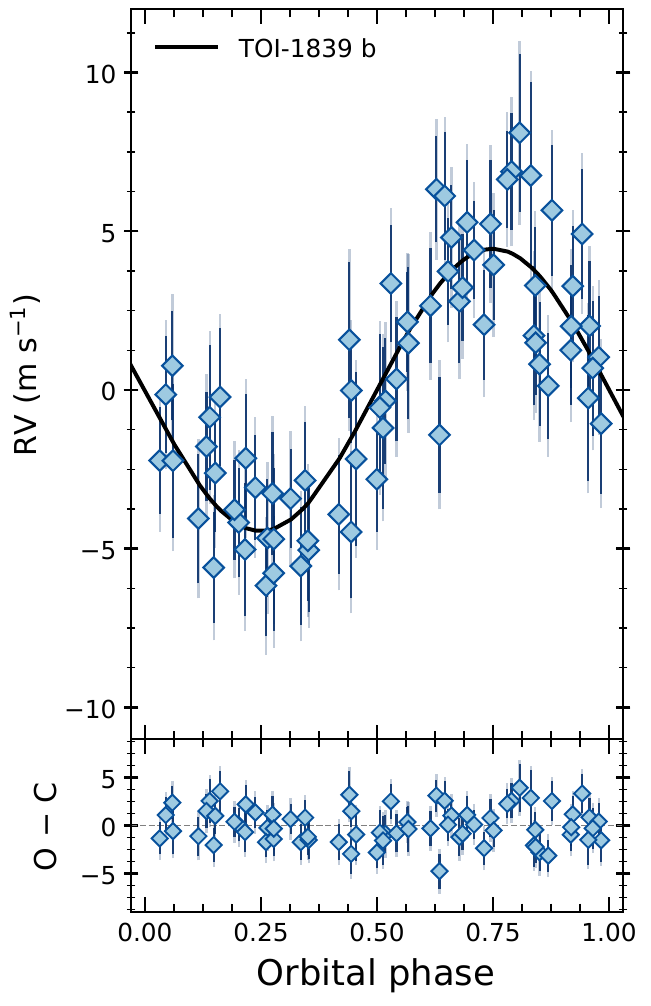}
    \includegraphics[width=0.229\textwidth]{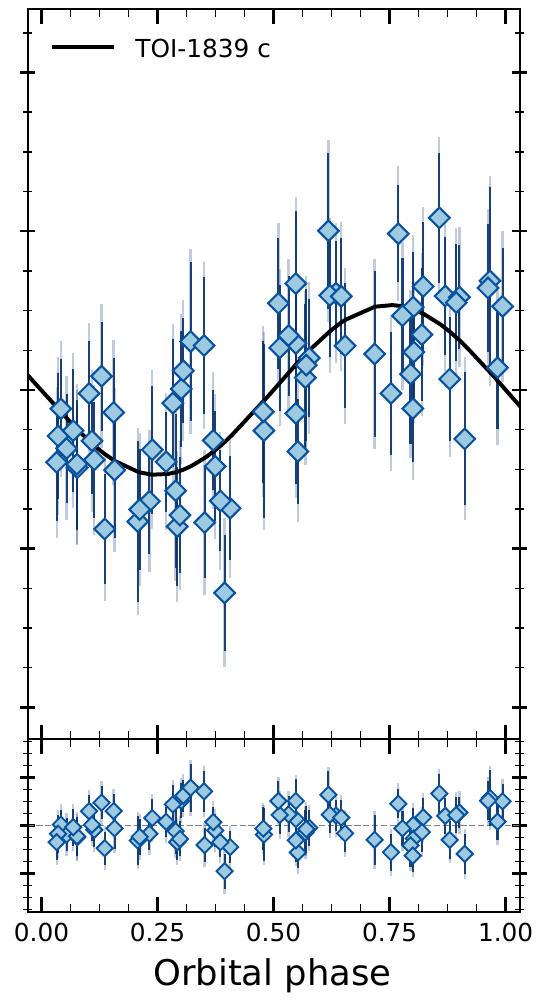}
    \includegraphics[width=0.229\textwidth]{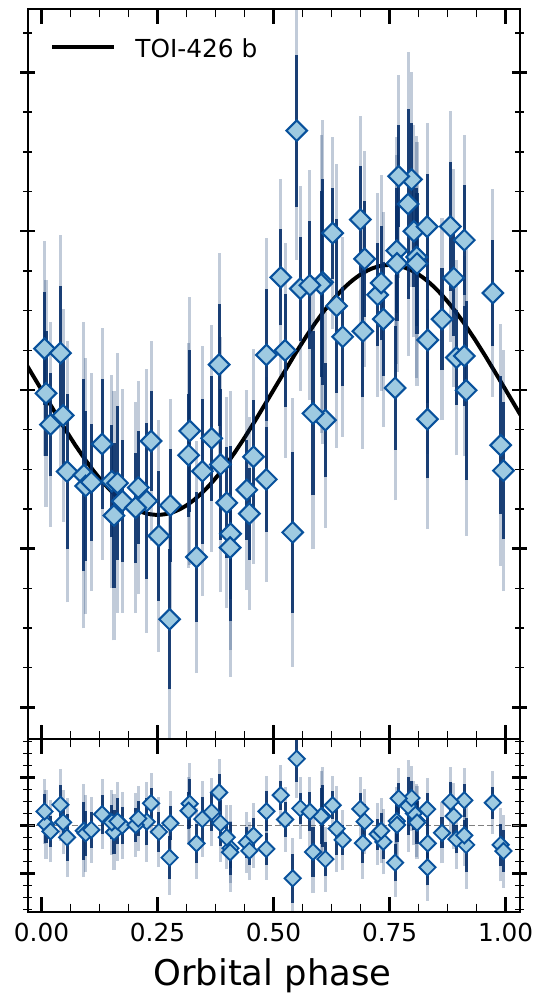}
    \includegraphics[width=0.25535\textwidth]{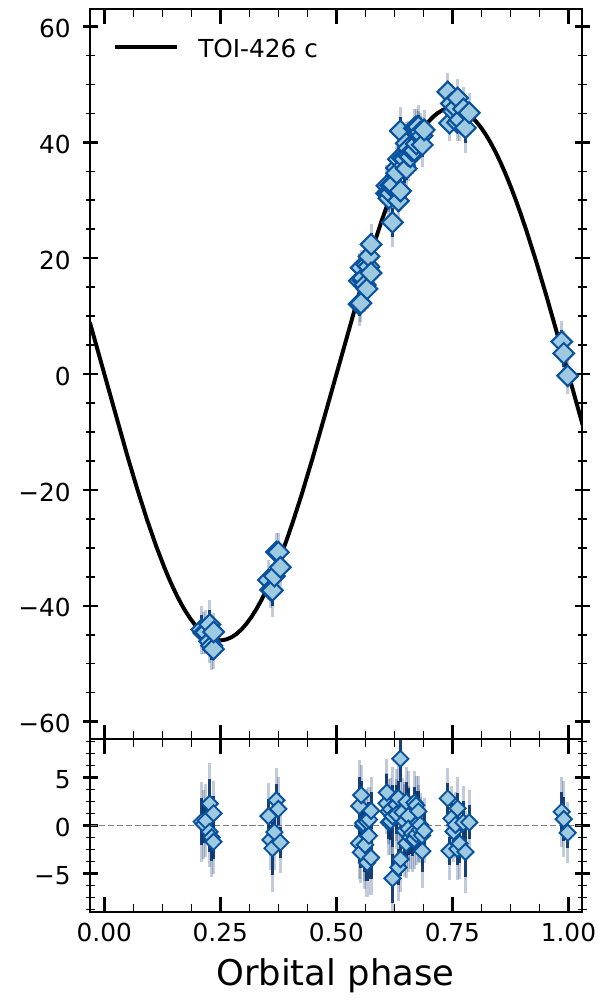}
    
    \caption{HARPS RVs induced by TOI-1839 b, TOI-1839 c, TOI-426 b, and TOI-426 c, phase-folded at their respective orbital periods. The blue error bars represent the instrumental uncertainties, and the light blue error bars represent the total uncertainties obtained by quadratically adding the corresponding jitter terms. The black solid lines indicate the median posterior Keplerian models.}
    \label{fig:RV_phase}
\end{figure*}

We performed model-comparison analyses for TOI-426 using both GP frameworks, exploring models with circular and eccentric Keplerian orbits for TOI-426~b, as well as models including an additional Keplerian signal. In these analyses, we used different combinations of the spectroscopic indicators H$\alpha$, $S$ index, FWHM, and BIS to inform the stellar activity modelling. Model comparisons based on $\Delta\ln\mathcal{Z}$ for the ST-GP models and on $\Delta$\mgic\ for the multi-GP models consistently support the planetary interpretation of TOI-426.01 (hereafter TOI-426~b), favouring a circular-orbit solution with a Doppler semi-amplitude detected at the $\sim3$--$4\sigma$ level. In addition, the multi-GP framework supports the presence of an additional long-period massive planet, TOI-426~c, associated with the high-amplitude RV signal identified in App.~\ref{sec:frequency}.

Figure~\ref{fig:posterior_426} (upper panels) shows the posterior distributions of $K$, $\lambda_{\rm e}$, $\lambda_{\rm P}$, and $P_{\rm GP}$ obtained from the ST-GP regressions using different spectroscopic indicators for the one-planet circular model. The Doppler semi-amplitudes of TOI-426~b are detected with significances of $3.0$--$3.2\sigma$ and are mutually consistent across the different activity indicators. We repeated this analysis including either a quadratic trend or an additional Keplerian signal in the RV model. For the latter, we adopted wide uniform priors and assumed a circular orbit. Both extensions yield $\Delta \ln \mathcal{Z} < 0$ relative to the one-planet model, indicating that the Bayesian evidence supports neither a quadratic trend nor an additional planet. We note, however, that the quadratic coefficient and the semi-amplitude of the additional Keplerian both differ from zero at the $\sim2\sigma$ level, with the latter showing a period posterior peaking at $\sim230$\,d. The $\sim2\sigma$ deviations from zero suggest that both extensions account for part of the observed long-term RV variability. However, the flexibility of the GP is sufficient to reproduce this variability without an additional component. The resulting degeneracy favours the simpler one-planet model.

We further investigated the origin of the long-term RV signal using the multi-GP framework. As a baseline, we first considered three RV-only 1D-GP models: one with no additional long-term component, one with a quadratic trend, and one with an additional circular Keplerian signal. The three configurations yield statistically indistinguishable \mgic\ values, with all differences below 10 \citep{mgic}, consistent with the degeneracy identified in the ST-GP analysis. We then explored the same three configurations using 2D-GP models in which the RVs are jointly modelled with one photometric-like activity indicator at a time ($S$ index, H$\alpha$, or FWHM) through a common latent process. This approach allowed us to test whether the long-term variability is shared between the RVs and the activity indicator. A signal present in both time series would favour a stellar activity origin, whereas a signal present only in the RVs would support a dynamical interpretation. As shown in Table~\ref{tab:mgic_toi426}, all 2D-GP models including an additional Keplerian signal are preferred over the quadratic-trend and no-trend alternatives, with all \mgic\ differences exceeding the adopted threshold of 10 (e.g. 58 and 178, respectively, for the RV+FWHM configuration). The \mgic\ comparison across dimensionalities further shows that all 2D-GP models including an additional Keplerian signal are preferred over any of the RV-only 1D-GP models, with \mgic\ values at least 34 lower than that of the best-performing 1D-GP model. The improvement over the 1D-GP models is also reflected in Fig.~\ref{fig:posterior_426}, where the Doppler semi-amplitude of TOI-426~b is more precisely constrained by the 2D-GP models, consistent with previous findings for activity-dominated time series \citep[e.g.][]{2015MNRAS.452.2269R,Barragan2023}. These results indicate that the long-term RV signal is neither captured by the shared activity process nor adequately described by a quadratic drift, supporting a planetary origin for the signal. We therefore interpret it as a second planet in the system, TOI-426~c.

Next we explored the addition of the BIS indicator to each RV+photometric-like indicator combination. In the multi-GP framework, BIS is an RV-like diagnostic that depends on both the activity process and its time derivative and is therefore expected to provide complementary constraints on the stellar RV signal \citep[e.g.][]{2012MNRAS.419.3147A,Barragan2023}. The resulting 3D-GP models are preferred over their corresponding 2D-GP models. Adding BIS lowers \mgic\ by 19 for the RV+FWHM combination and by 51 for both the RV+H$\alpha$ and RV+$S$ index combinations, with all three improvements exceeding the adopted threshold of 10. This improvement is also evident in Fig.~\ref{fig:posterior_426}, where adding BIS increases the precision of the inferred semi-amplitudes and largely removes the secondary mode in $\lambda_{\rm e}$ present in the 1D-GP and 2D-GP posteriors. Within the 3D-GP framework, the three indicator combinations yield nearly identical \mgic\ values, ranging from $-484$ to $-483$.

\begin{figure*}
    \centering
    \includegraphics[width=0.9\columnwidth]{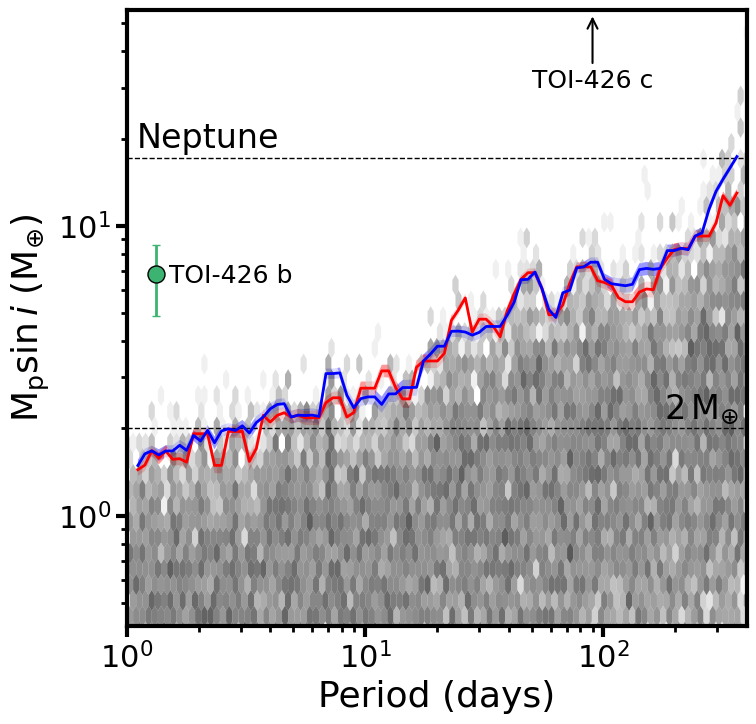}
    \includegraphics[width=0.9024\columnwidth]{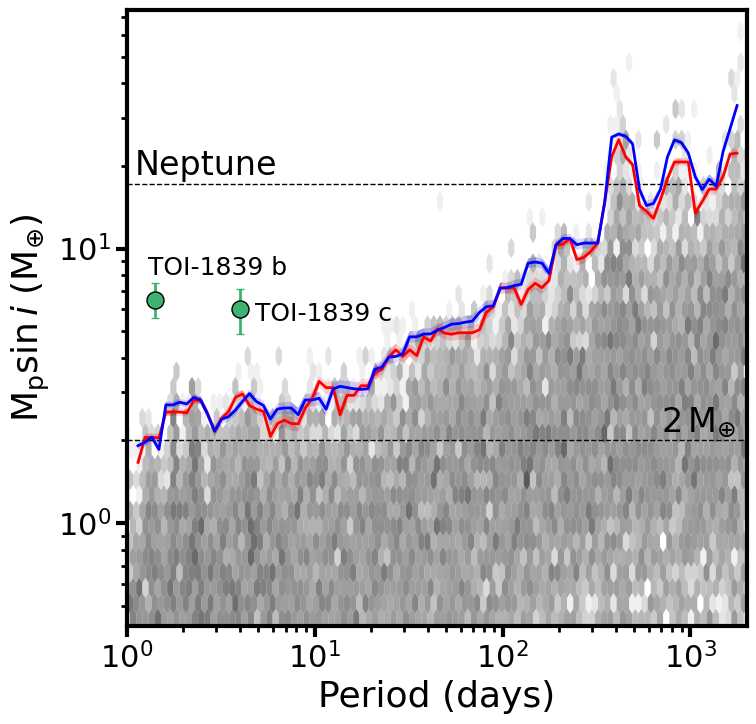}
    \caption{Hexbin plot of the posterior samples from one-planet fits to the RV residuals of TOI-426 (left) and TOI-1839 (right). The blue curves show the $3\sigma$ sensitivity limits, while the red curves show the corresponding limits restricted to samples with $e<0.1$.}
    \label{fig:TOI-426_Mass_Detection_lim}
\end{figure*}

We also explored the orbital shape of TOI-426~c by allowing for an eccentric orbit in all three 3D-GP configurations. The eccentric solutions yield \mgic\ values 2--3 higher than their circular counterparts and are therefore slightly disfavoured (Table~\ref{tab:mgic_toi426}). Because their additional free parameters do not meaningfully improve the RV likelihood, we retained the simpler circular parametrisation as our adopted solution. This choice, however, should not be interpreted as evidence that the orbit of TOI-426~c is intrinsically circular. Because both extrema of the RV signal are well sampled, allowing the eccentricity to vary leaves the Doppler semi-amplitudes of both planets essentially unchanged, with the eccentric fit yielding $K_c = 45.5^{+1.7}_{-1.6}$\,m\,s$^{-1}$, consistent with the circular value of $45.9 \pm 1.4$\,m\,s$^{-1}$. However, the sparse sampling of some orbital phases broadens the range of compatible orbits and introduces clear covariances among $P_c$, $T_{0,c}$, $\omega_c$, and $e_c$ (the eccentric fit yielded $e_c = 0.10^{+0.09}_{-0.07}$, $P_c = 256^{+33}_{-22}$\,d, and $T_{0,c} = 9781^{+18}_{-27}$ RJD, while $\omega_c$ remains poorly constrained). We therefore caution that the reported uncertainties on the ephemeris are conditional on the adopted circular parametrisation and do not capture the full range permitted by the eccentric solutions. Additional RV observations are needed to further constrain the orbital shape of TOI-426~c.

Overall, given that the three circular 3D-GP models yield statistically indistinguishable \mgic\ values, we adopted the model informed by H$\alpha$ and BIS as our final baseline because it provides slightly higher detection significances for both Keplerian signals, with $K_b/\sigma_{K_b} \approx 3.7$ and $K_c/\sigma_{K_c} \approx 33.1$. The full RV model and the phase-folded RV curves are shown in Figs.~\ref{fig:426_RV_full} and \ref{fig:RV_phase}, respectively, and the corresponding posterior and derived parameters are listed in Table~\ref{tab:toi426_toi1839_priors_posteriors}.

\subsubsection{Results on TOI-1839}
\label{sec:TOI-1839}

We performed similar model-comparison analyses for TOI-1839 using both GP frameworks, exploring models with circular and eccentric orbits for the transiting planets, as well as models including additional non-transiting companions. Model comparisons based on $\Delta\ln\mathcal{Z}$ for the ST-GP models and on $\Delta$\mgic\ for the multi-GP models consistently favour the simplest configuration: two Keplerians in circular orbits associated with TOI-1839.01 and TOI-1839.02 (hereafter TOI-1839\,b and c).

Figure~\ref{fig:posterior_1839} (upper panels) shows the posterior distributions of $K$, $\lambda_{\rm e}$, $\lambda_{\rm P}$, and $P_{\rm GP}$ obtained from the ST-GP regressions using different spectroscopic indicators. The recovered Doppler semi-amplitudes are statistically significant ($9.8\sigma$ for TOI-1839\,b and $5.3\sigma$ for TOI-1839\,c) and consistent across all activity indicators. We note that although the $S$ index does not significantly constrain the stellar signal, the inferred semi-amplitudes are consistent with those obtained using FWHM and H$\alpha$. Figure~\ref{fig:posterior_1839} (lower panels) also shows the posteriors from the 2D-GP regressions using the same spectroscopic indicators. Among all the tested models, the largest $t$-values are 0.32 for TOI-1839\,b and 0.64 for TOI-1839\,c, confirming that the inferred semi-amplitudes are insensitive to the choice of GP framework and activity indicator. The 2D-GP configurations informed by H$\alpha$ and FWHM yield statistically indistinguishable \mgic\ values, whereas the corresponding $S$ index configuration is less favoured, consistent with the weaker constraints it provides on the activity hyperparameters (Fig.~\ref{fig:posterior_1839} and Table~\ref{tab:mgic_toi1839}). Overall, although the multi-GP can in principle provide tighter constraints by exploiting the covariance between the RVs and the activity indicators, it yields no measurable improvement over the ST-GP results. This might be due to the large RV gaps and the low S/N of the activity signals \citep[][]{2024MNRAS.531.4275B}.

Given these results, we adopted the simpler ST-GP framework, with GP timescales informed by the H$\alpha$ indicator, as the final baseline model. The phase-folded RV curves of TOI-1839\,b and TOI-1839\,c are shown in Fig.~\ref{fig:RV_phase}, and the posterior parameters are listed in Table~\ref{tab:toi426_toi1839_priors_posteriors}. The full model is shown in Fig.~\ref{fig:1839_RV_full}.

\subsection{Sensitivity limits and constraints on additional planets}
\label{sec:sensitivity_limits}

We derived RV sensitivity limits following \citet{Standing2022} and \citet{Standing2026}. After subtracting the Keplerian and activity signals from the RV time series of TOI-426 and TOI-1839, we analysed the residuals with \texttt{kima} \citep{Faria2018} to search for additional planetary signals. Once it was verified that no significant signals remained in the residuals, we ran \texttt{kima} with the number of planetary signals ($N_{\mathrm{p}}$) fixed to one. The resulting posterior samples are consistent with the residual data, but correspond to signals that are not significantly detected.

In total, more than 60\,000 posterior samples were obtained for both TOI-426 and TOI-1839. A $3\sigma$ upper limit was then computed in bins of orbital period, yielding the sensitivity limits shown by the blue and red curves in Fig.~\ref{fig:TOI-426_Mass_Detection_lim}. For low-eccentricity orbits, the available data are sensitive at the $3\sigma$ level to planetary signals lying above the red limits, as is the case for TOI-426\,b and c, and TOI-1839\,b and c. From Fig.~\ref{fig:TOI-426_Mass_Detection_lim}, we conclude that, for low-eccentricity orbits in both systems, we can rule out the presence of planets with minimum masses $M_{\mathrm{p}}\sin i \gtrsim 2\,\mathrm{M_{\oplus}}$ at $P_{\rm orb} \lesssim 1.5$\,d, $M_{\mathrm{p}}\sin i \gtrsim 3\,\mathrm{M_{\oplus}}$ at $P_{\rm orb} \lesssim 10$\,d, and $M_{\mathrm{p}}\sin i \gtrsim 10\,\mathrm{M_{\oplus}}$ at $P_{\rm orb} \lesssim 300$\,d. We note that although the blue curves include samples with higher eccentricities, the projected limits do not capture the full dependence of detectability on eccentricity, the argument of periastron, and the time of periastron passage. Highly eccentric companions whose periastron passages coincide with the large gaps in the RV time series may remain undetected.

\begin{figure*}
    \centering
    \includegraphics[width=0.98\columnwidth]{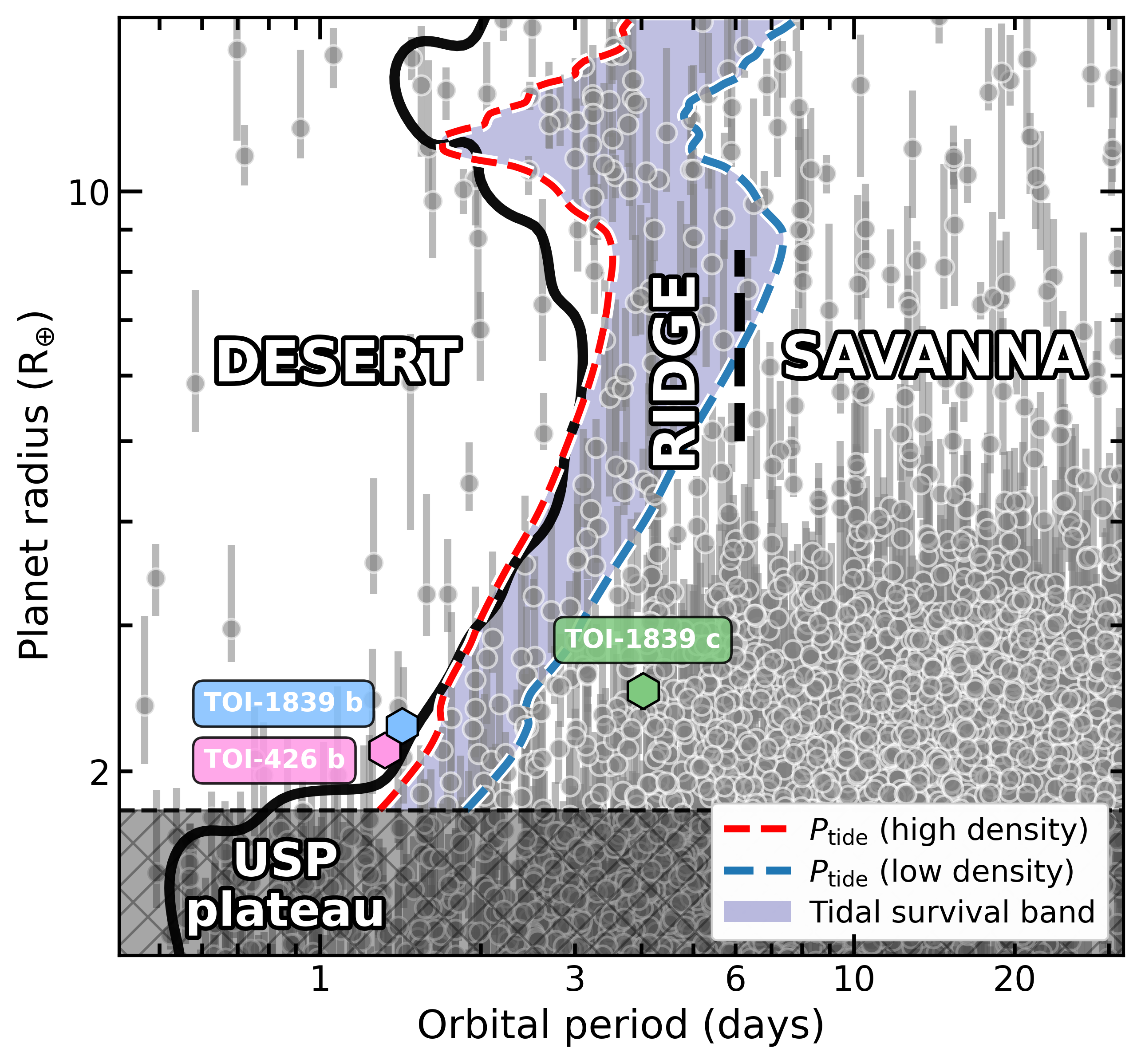}
    \includegraphics[width=0.996\columnwidth]{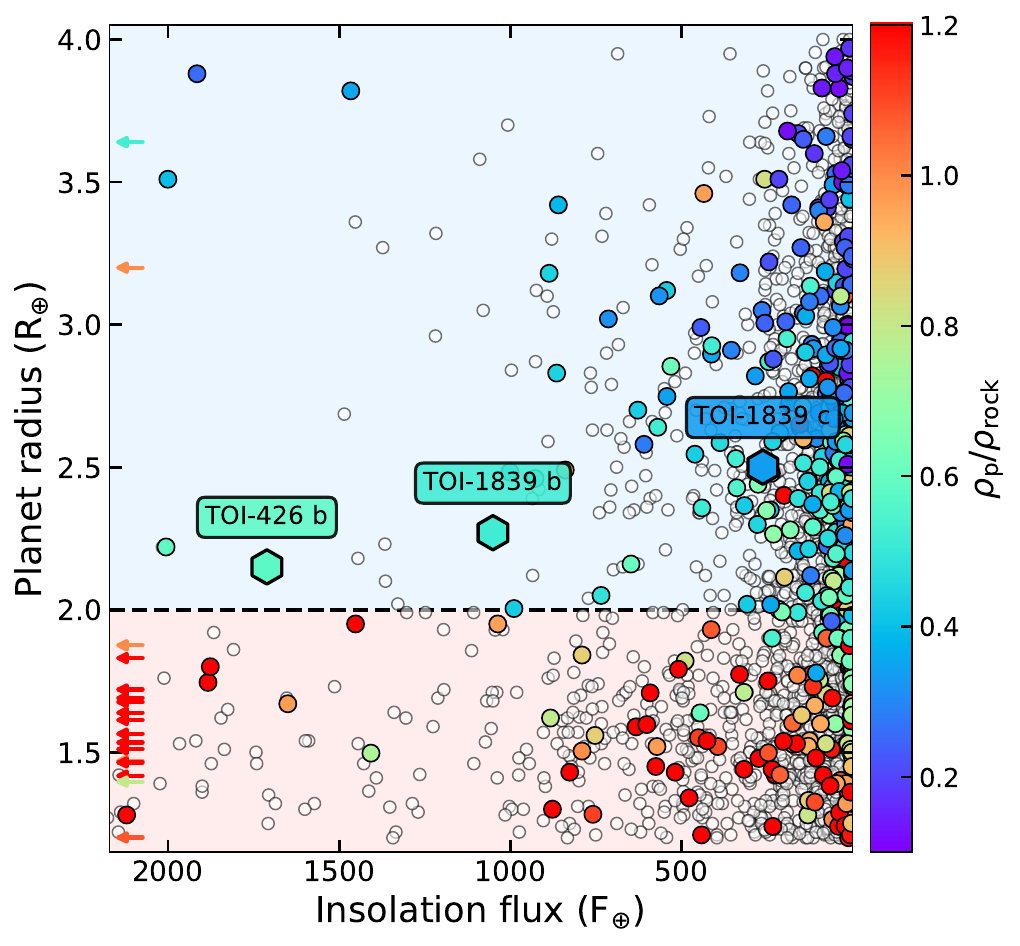}
    \caption{Left panel: Planet radius versus orbital period for the \textit{Kepler} DR25 catalogue, used by \citet{2024A&A...689A.250C} to delineate the bias-corrected boundaries of the Neptunian desert, ridge, and savanna. The red and blue dashed lines correspond to the tidal survival boundaries following HEM \citep{2026A&A...709L..17C}. The region between these lines corresponds to the tidal survival band, within which HEM outcomes are expected to be deposited. Right panel: Planet radius versus insolation flux for all known sub-Neptunes catalogued in the NASA Exoplanet Archive. The left panel was created with \href{https://github.com/castro-gzlz/nep-des}{\texttt{nep-des}}.}
    \label{fig:nep_des}
\end{figure*}

\section{Discussion}
\label{sec:discussion}

Close-in sub-Neptunes with $P_{\rm orb} \lesssim 1.5\,\mathrm{d}$ are rare in exoplanet catalogues, making TOI-426\,b and TOI-1839\,b valuable laboratories for investigating planetary evolution under extreme conditions. In this section, we place these planets in the context of the Neptunian desert, examine the role of outer companions near the lower desert boundary, and investigate their volatile content through interior-structure and atmospheric-escape modelling.

\subsection{Sub-Neptunes at the edge of the Neptunian desert}
\label{sec:desert-edge}

In Fig.~\ref{fig:nep_des} (left panel), we place TOI-426\,b, TOI-1839\,b, and TOI-1839\,c in the period--radius plane, together with the geometry of the Neptunian desert derived by \citet{2024A&A...689A.250C}. We note that both TOI-426\,b and TOI-1839\,b lie close to the lower boundary. \citet{2026A&A...709L..17C} recently showed that the slope of this boundary and the overall desert morphology are naturally reproduced by the tidal survival boundary \citep[e.g.][]{Roche1849} following HEM \citep[e.g.][]{Ford2006}, $P_{\rm tide}$, supporting this mechanism as a primary driver of the desert shape \citep[e.g.][]{Matsakos2016}. In that work, $P_{\rm tide}$ was computed across the observed mass--radius distribution assuming a single tidal encounter parameter, yielding a tidal survival band within which HEM outcomes are expected to be effectively deposited. This band closely follows the extension of the Neptunian ridge \citep{2024A&A...689A.250C} and the hot-Jupiter pile-up \citep{2007ARA&A..45..397U}, suggesting that these occurrence overdensities are likely populated, at least in large part, by the accumulation of HEM products \citep[e.g.][]{Dawson2018,Correia2020,Bourrier2023,Bourrier2025}. In the sub-Neptune regime, however, no analogous overdensity has been identified, so additional observables such as the prevalence of outer giant planets may provide some of the clearest evidence of an underlying HEM contribution near the survival band. We explore this possibility further in Sect.~\ref{sec:outer_companions}.

\subsection{Volatile-rich sub-Neptunes under extreme irradiation}
\label{sec:volatile-rich}

In Fig.~\ref{fig:nep_des} (right panel), we contextualise TOI-426\,b, TOI-1839\,b, and TOI-1839\,c in an insolation--radius diagram with the current sample of known small planets. Sub-Neptune planets are typically found at $F \lesssim 300\,\mathrm{F}_{\oplus}$, whereas super-Earths span a much wider range of insolations. Notably, the combination of their short orbits and the relatively early spectral types of their hosts places TOI-426\,b and TOI-1839\,b among the most highly irradiated sub-Neptunes known to date. In particular, TOI-426\,b receives the second-highest insolation ($F = 1710 \pm 110\,\mathrm{F}_{\oplus}$) among planets in the upper mode of the bimodal radius distribution (i.e. $2\,\mathrm{R}_{\oplus} < R_{\mathrm{p}} < 3\,\mathrm{R}_{\oplus}$), surpassed only by TOI-1408~c \citep[$F \approx 2000\,\mathrm{F}_{\oplus}$;][]{2024ApJ...971L..28K}. With this increased sample, we explore whether the typically volatile-rich sub-Neptune population remains volatile-rich when subject to such extreme irradiation. In this diagram, planets are colour-coded according to their bulk densities, normalised to the density expected for a planet with a 100\,\% silicate composition \citep{2019PNAS..116.9723Z}. Canonical mildly irradiated sub-Neptunes (e.g. $R_{\mathrm{p}} > 2\,\mathrm{R}_{\oplus}$; $F \lesssim 300\,\mathrm{F}_{\oplus}$) typically exhibit $\rho_{\mathrm{p}}/\rho_{\mathrm{rock}} < 1$, whereas super-Earths (e.g. $R_{\mathrm{p}} < 2\,\mathrm{R}_{\oplus}$) generally show $\rho_{\mathrm{p}}/\rho_{\mathrm{rock}} > 1$. Notably, despite their extreme irradiation levels, TOI-426\,b and TOI-1839\,b require the presence of volatile elements to explain their densities ($\rho_{\mathrm{p}}/\rho_{\mathrm{rock}} < 1$). 

Primordial \ce{H2}/He envelopes of sub-Neptunes are expected to be efficiently eroded under high irradiation \citep[e.g.][]{2017ApJ...847...29O}. The existence of highly irradiated volatile-rich sub-Neptunes such as TOI-426\,b and TOI-1839\,b may therefore be explained by scenarios in which the volatiles are not directly exposed to stellar irradiation \citep[i.e. buried within the planetary interiors; e.g.][]{2024NatAs...8.1399L}, or are more resistant to it \citep[i.e. atmospheres composed of high mean molecular weight volatiles; e.g.][]{2020A&A...638A..41T}. In Sect.~\ref{sec:volatiles}, we further explore the internal structures of TOI-426\,b, TOI-1839\,b, and TOI-1839\,c, as well as the impact of atmospheric escape under different scenarios.

\begin{figure*}
    \centering
    \includegraphics[width=1\textwidth]{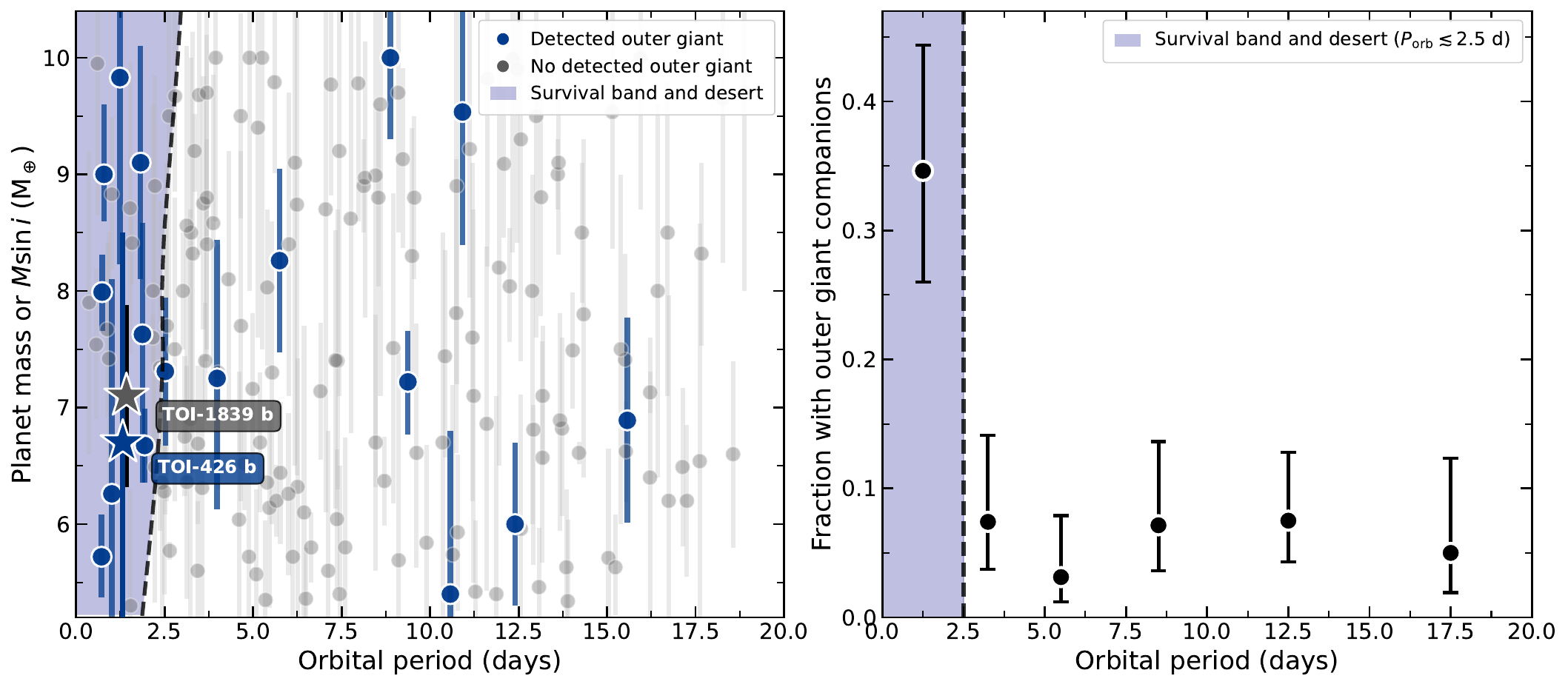}
    \caption{Observed incidence of outer giant companions to short-period sub-Neptunes with masses or minimum masses up to 10~$\rm M_{\oplus}$. Left: Planet mass, or minimum mass, versus orbital period for the innermost sub-Neptune in each system. Systems hosting outer giant companions ($M_{\rm p} \geq 0.3\,\rm M_{\rm J}$, $P_{\rm orb} > 100$~d) are highlighted. The shaded region marks the tidal survival band expected after HEM \citep{2026A&A...709L..17C}. Right: Observed fraction of systems with outer giant companions as a function of the inner sub-Neptune period. The shaded region corresponds to a simplified approximation of the tidal survival band at $P_{\rm orb} \lesssim 2.5$~d.}
    \label{fig:outer_giant_companions}
\end{figure*}

\subsection{Outer giant companions to sub-Neptunes near the edge of the Neptunian desert}
\label{sec:outer_companions}

The orbital architecture of the TOI-426 system, namely a short-period sub-Neptune close to the tidal survival band expected after HEM \citep{2026A&A...709L..17C}, accompanied by a long-period, Jupiter-mass planet, makes TOI-426\,b a compelling candidate for having undergone HEM. This mechanism has been widely invoked to explain the orbits of short-period giant planets \citep[e.g.][]{2011ApJ...735..109W}, and it has also been shown to operate effectively in the small planet regime. For example, \citet{2019AJ....157..180P} show that sub-Neptunes initially formed at periods of 5--10 days can undergo HEM through secular chaos over a wide range of architectures, including configurations with companions of masses $\gtrsim$ 10~$\mathrm{M_{\oplus}}$ on orbits ranging from periods of tens of days to au-scale separations. Notably, \citet{Doyle2025} find that close-in sub-Neptunes near the lower boundary of the desert frequently have companions. Here, we focus on the population of close-in sub-Neptunes with long-period giant planets, such as TOI-426~b, and discuss the implications for the HEM interpretation of the desert.

In Fig.~\ref{fig:outer_giant_companions}, we show the mass--period diagram of sub-Neptunes with masses or minimum masses ranging from $5.2\,\rm M_{\oplus}$ (roughly corresponding to $1.8\,\rm R_{\oplus}$) to 10~$\rm M_{\oplus}$ and mass precisions better than 33\%, including TOI-426\,b and TOI-1839\,b. To avoid overweighting multi-planet systems, we included only the innermost planet in each system. We highlight systems hosting long-period ($P_{\rm orb} > 100$ d) giant planets ($M_{\rm p} \geq 0.3\,\rm M_{\rm J}$). Within this framework, we find a striking concentration of outer companions among the shortest-period sub-Neptunes, specifically those located within the desert and survival band: 9 of 25 planets in this narrow region have an outer giant companion, compared to only 9 of 148 outside it, corresponding to detected fractions of $36.0\%$ and $6.1\%$, respectively. This difference is highly significant (Fisher exact test $p = 1.4\times10^{-4}$; permutation test $p = 1.2\times10^{-4}$). To test the robustness of this trend, we performed a systematic exploration of the parameter space, varying the mass precision thresholds (33\%, 25\%, and 20\%), the minimum companion mass ($> 0.3$, 0.5, 0.7, 1.0, and 1.5~$\rm M_{\rm J}$), and the minimum orbital period ($> 0$, 50, 100, 150, 200, and 300~d) used to define giant companions. Across all 90 tested configurations, the enrichment of outer giant companions within the tidal survival band remains statistically significant (Fisher exact test $p < 0.05$ in 100\% of cases), with p-values spanning $1.4\times10^{-4}$ to $2.8\times10^{-2}$. The difference in detection rates is consistently positive, corresponding to typical companion fractions of $\sim24\%$--$40\%$ inside the band compared to $\sim5\%$--$9\%$ outside it. This shows that the observed trend is not driven by the adopted companion definition. We also verified that the companion excess remains statistically significant when extending the inner-planet mass range up to $20\,\rm M_{\oplus}$, approximately spanning the broader sub-Neptune domain, although the limited number of planets at the high-mass end prevents us from assessing whether the trend holds independently in that regime. As a complementary check, we repeated the analysis using a radius-selected sub-Neptune sample with measured masses ($2.0 \leq R_{\rm p} \leq 3.5\,\rm R_{\oplus}$), broadly tracing the volatile-rich sub-Neptune regime and confirming that the trend is not driven by the inclusion of smaller rocky planets. Taken together, these tests show that the excess of outer giant companions is robust to the adopted sample definition and is not an artefact of the overlap between rocky and volatile-rich planets in mass. They therefore strengthen the evidence of this trend among low-mass, volatile-rich sub-Neptunes near the lower edge of the desert.

We note that the construction of the sample mitigates several key observational biases. The requirement that the inner sub-Neptunes have well-constrained masses inherently selects systems with extensive RV data sets. In particular, longer-period planets require more follow-up on average to reach comparable mass precision, favouring the detection of outer companions, contrary to the observed trend. To obtain an initial assessment of the impact of heterogeneous follow-up, we compared the RV data sets of the 9 systems hosting outer giant companions within the tidal survival band to those of 30 randomly selected systems outside the band without detected companions. Both groups exhibit comparable data sets, with several tens to a few hundred RVs and baselines typically exceeding 100--200 days and often extending beyond 500--1000 days. Moreover, only 4 of the 30 comparison systems ($13\%$) show long-term RV trends consistent with outer companions, well below the $\sim36\%$ fraction observed within the tidal survival band. This suggests that the observed excess is unlikely to be driven by differences in observing baseline. However, a dedicated occurrence-rate study will be required to robustly quantify the underlying population-level trend.

Overall, the large $\sim$35\% fraction of low-mass sub-Neptunes on the shortest-period orbits with outer long-period giant companions is consistent with an important contribution from HEM in populating this region, as discussed by \citet{Matsakos2016} and \citet{2026A&A...709L..17C}. The HEM interpretation may also extend to a subset of the ultra-short-period rocky planet population, which may represent the end states of tidally disrupted sub-Neptunes \citep[e.g.][]{2019AJ....157..180P,2021PSJ.....2..152A,2025A&A...699A.344C}. Further dedicated simulations, informed by the properties of these systems, will be required to determine which HEM pathways may preferentially shape this population. Notably, when constrained, the outer giant companions often exhibit moderate-to-high eccentricities (typically $e \sim 0.3$--0.6, reaching up to $e \sim 0.85$), further supporting high-eccentricity excitation. Whether TOI-426~c shares this characteristic remains uncertain. As discussed in Sect.~\ref{sec:TOI-426}, the current data do not support the additional complexity of an eccentric model, nor do they establish that its orbit is intrinsically circular. Additional RV monitoring is required to better constrain its orbital shape.

TOI-1839, in contrast, hosts TOI-1839~b near the survival band but shows no evidence of an outer giant companion. As discussed in Sect.~\ref{sec:TOI-1839}, the current RVs do not reveal a significant long-term trend. For low-eccentricity orbits, the sensitivity limits shown in Fig.~\ref{fig:TOI-426_Mass_Detection_lim} rule out giant companions with periods up to approximately 2000~d. However, the large gaps in the RV time series may still allow highly eccentric companions whose periastron passages fall within unsampled intervals, as well as companions with longer periods. Additional well-sampled RV observations are therefore required to robustly constrain the architecture of this system. We note that the short orbital separation and low mass of TOI-1839~c make it unlikely to act as a HEM perturber for TOI-1839~b \citep[e.g.][]{2019AJ....157..180P}. However, alternative pathways such as low-eccentricity or obliquity-driven migration may operate through dynamical interactions between close-in planets such as TOI-1839~b and TOI-1839~c \citep{2019MNRAS.488.3568P,2020ApJ...905...71M}. Overall, whether TOI-1839~b reached the lower edge of the desert via HEM triggered by an as-yet undetected companion, through interactions with TOI-1839~c, or alternatively through disk-driven migration or in situ formation remains an open question that will require improved observational constraints together with dedicated theoretical and numerical studies.

\subsection{Volatile content of the three transiting planets}
\label{sec:volatiles}

\begin{figure}
    \centering
    \includegraphics[width=0.98\columnwidth]{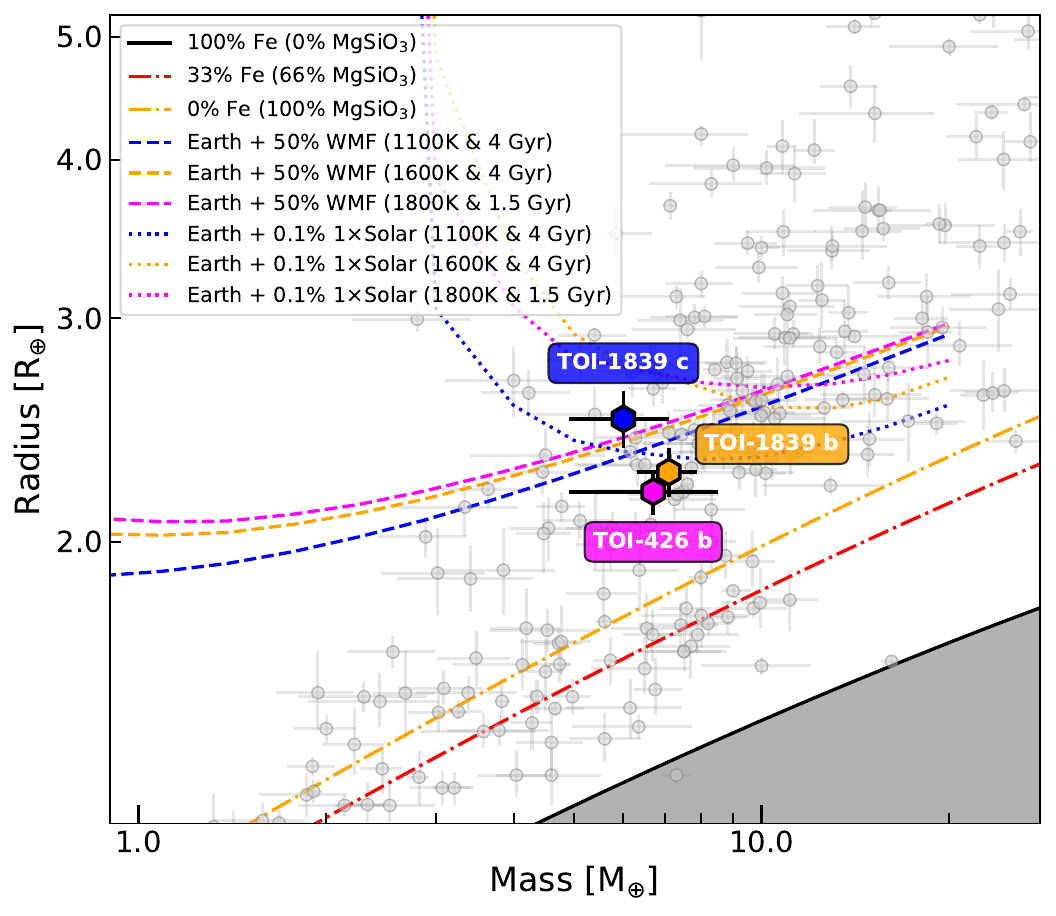}
    \caption{Mass--radius diagram of planets with precise masses and radii from the NASA Exoplanet Archive. Curves show rocky models \citep{2019PNAS..116.9723Z}, steam-world models \citep{2025ApJ...988..186A,aguichine_2025_17811236}, and gas-dwarf models \citep{2025ApJ...989...28T}. This plot was created with \href{https://github.com/castro-gzlz/mr-plotter}{\texttt{mr-plotter}} \citep{2023A&A...675A..52C}.}
    \label{fig:mr-plotter}
\end{figure}

As shown in the mass--radius diagram in Fig.~\ref{fig:mr-plotter}, the masses and radii of TOI-426\,b, TOI-1839\,b, and TOI-1839\,c require the presence of a non-negligible volatile component, as their densities are inconsistent with purely rocky compositions at the $>3\sigma$ level. Among them, TOI-1839\,c is compatible with either a substantial water-rich layer ($\gtrsim 50\%$ by mass) or a thin \ce{H2}/He envelope ($\gtrsim 0.1\%$ by mass). Despite being subject to higher irradiation, TOI-1839\,b exhibits a smaller radius than its outer companion, suggesting a lower volatile content, while TOI-426\,b, exposed to even more extreme irradiation, is expected to retain an even smaller volatile reservoir.

To further constrain the nature of these volatiles, we modelled the internal structures and atmospheric mass-loss rates of the three planets under different assumptions, as detailed in Appendix~\ref{sec:interior_escape}. We considered (a) \ce{H2}/He-dominated envelopes spanning a range of metallicities ($1\times$ and $50\times$ solar) and (b) high mean molecular weight envelopes dominated by \ce{H2O}. For \ce{H2}/He envelopes, we infer short mass-loss timescales of $\sim1$--100\,Myr, indicating that such primordial atmospheres would be efficiently removed over the system lifetimes. In contrast, \ce{H2O}-dominated atmospheres yield much longer timescales of $\sim100$--1000\,Gyr, allowing them to be retained.

These findings disfavour a classical gas-dwarf scenario for TOI-426\,b, TOI-1839\,b, and TOI-1839\,c, even under relatively recent migration scenarios, which would require finely tuned timing to preserve \ce{H2}/He envelopes. Instead, they point towards high mean molecular weight atmospheres as a natural explanation for the survival of volatile-rich planets across a wide range of irradiations, from the extreme environments of TOI-426\,b and TOI-1839\,b to the more moderate conditions of TOI-1839\,c.

\section{Conclusions}
\label{sec:conclusions}

We have confirmed and characterised the planetary systems TOI-426 and TOI-1839, which host two sub-Neptunes near the lower edge of the Neptunian desert: TOI-426~b ($P_{\rm orb} \approx 1.32$ d, $R_{\rm p} = 2.19 \pm 0.09$~$\rm R_{\oplus}$) and TOI-1839~b ($P_{\rm orb} \approx 1.42$ d, $R_{\rm p} = 2.27 \pm 0.10$~$\rm R_{\oplus}$). We measured their radii and masses using TESS and HARPS data, which also allowed us to detect the companions TOI-1839~c ($P_{\rm orb} \approx 4.02$~d, $R_{\rm p} = 2.50 \pm 0.13$~$\rm R_{\oplus}$) and TOI-426~c (assuming $e = 0$, $P_{\rm orb} = 235.8^{+9.2}_{-8.5}$ d, $M_{\rm p}\sin i = 444^{+18}_{-17}$~$\rm M_{\oplus}$).

We performed RV analyses using both the ST-GP and multi-GP frameworks to account for stellar activity. For the active and relatively young star TOI-426, the 2D-GP and 3D-GP configurations provide an improved description of the data compared with both the ST-GP and RV-only 1D-GP models, yielding a more precise mass for TOI-426~b and enabling the confirmation of the long-period planet TOI-426~c. This highlights the ability of the multi-GP framework to model complex stellar signals and to disentangle activity-driven trends from long-period planetary signals that may otherwise be absorbed in simpler GP models.

The orbital architecture of TOI-426, namely a short-period sub-Neptune close to the tidal survival band expected after HEM accompanied by a long-period giant planet, makes it a compelling system for having undergone this process. In this context, we find a high incidence ($\approx$35\%) of outer giant companions to low-mass sub-Neptunes near the survival band, consistent with an important role for HEM in shaping the lower desert boundary. In contrast, TOI-1839 does not currently show evidence of an outer giant companion, although we cannot rule out its presence based on the existing RVs. This leaves open the possibility that TOI-1839~b may have reached its current orbit through HEM or through alternative pathways, including disk-driven migration or in situ formation, highlighting the need for additional observations and occurrence-rate studies to robustly constrain the system architectures near the lower desert edge.

The combination of their short orbital periods and the relatively early spectral types of their host stars places TOI-426~b and TOI-1839~b among the most highly irradiated sub-Neptunes known to date, making them prime targets for probing planetary evolution under extreme irradiation conditions. Notably, the density constraints indicate that both planets require a significant amount of volatile material. Assuming \ce{H2}/He-dominated atmospheres, we estimate mass-loss timescales of order 1--100\,Myr, making the classical gas-dwarf scenario unlikely. This conclusion extends to the less irradiated companion TOI-1839~c, for which a \ce{H2}/He envelope would also be efficiently removed on similarly short ($\sim$Myr) timescales. In contrast, assuming \ce{H2O}-dominated atmospheres, the mass-loss timescales increase to $\sim100$--1000\,Gyr, supporting a steam-world interpretation for all three planets. The survival of highly irradiated volatile-rich sub-Neptunes near the lower edge of the desert is difficult to reconcile with a classical picture in which this boundary is set solely by the evaporation of primordial \ce{H2}/He envelopes, and instead favours a scenario in which water-rich steam worlds can persist in this region. Future work will be required to assess the long-term stability of the lower edge of the desert against the evolution and erosion of steam-rich atmospheres.

\section*{Data availability}

The full versions of Tables D.6--D.11 are available at the \href{http://cdsweb.u-strasbg.fr/cgi-bin/qcat?J/A+A/}{CDS}.

\begin{acknowledgements}

We are particularly grateful to the referee for their thoughtful and constructive report, which greatly helped improve this work.

This work has been carried out within the framework of the NCCR PlanetS supported by the Swiss National Science Foundation under grant 51NF40\_205606.

This research was funded in part by the UKRI (Grants ST/X001121/1, EP/X027562/1).

This material is based upon work supported by NASA's Interdisciplinary Consortia for Astrobiology Research (NNH19ZDA001N-ICAR) under award number 80NSSC21K0597.

A.M. and P.F. acknowledge financial support from the Agencia Estatal de Investigaci\'on (AEI/10.13039/501100011033) of the Ministerio de Ciencia, Innovaci\'on y Universidades (MCIU) through project PID2023-149439NB-C44 (A.M.), co-funded by the European Regional Development Fund (ERDF) ``A way of making Europe'', and the Centre of Excellence ``Severo Ochoa'' award to the Instituto de Astrof\'isica de Andaluc\'ia (CEX2021-001131-S; P.F.). A.M. also acknowledges support from Generalitat Valenciana in the frame of the GenT Project ESGENT-CIESGT2025. P.F. is also funded by the European Union (ERC, THIRSTEE, 101164189). Views and opinions expressed are however those of the author(s) only and do not necessarily reflect those of the European Union or the European Research Council. Neither the European Union nor the granting authority can be held responsible for them.

V.~A. acknowledges support from Funda\c{c}\~ao para a Ci\^encia e Tecnologia (FCT) through national funds under the research grant UID/04434/2025 (DOI: 10.54499/UID/04434/2025) and through a work contract funded by the FCT Scientific Employment Stimulus programme (reference 2023.06055.CEECIND/CP2839/CT0005; DOI: 10.54499/2023.06055.CEECIND/CP2839/CT0005).

A.C.M.C. acknowledges support from the Swiss National Science Foundation under grant 51NF40\_205606.

F.P. acknowledges the Swiss National Science Foundation (SNSF) for supporting research with HARPS through SNSF grants 140649, 152721, 166227, 184618, and 215190.

This publication made use of \texttt{TESS-cont} \citep[][\url{https://github.com/castro-gzlz/tess-cont}]{2024A&A...691A.233C}, which also made use of \texttt{tpfplotter} \citep{2020A&A...635A.128A} and \texttt{TESS\_PRF} \citep{2022ascl.soft07008B}.

This paper made use of data collected by the TESS mission and made publicly available from the Mikulski Archive for Space Telescopes (MAST), operated by the Space Telescope Science Institute (STScI). Funding for the TESS mission is provided by NASA's Science Mission Directorate. We acknowledge the use of public TESS data from pipelines at the TESS Science Office and at the TESS Science Processing Operations Center. Resources supporting this work were provided by the NASA High-End Computing (HEC) Program through the NASA Advanced Supercomputing (NAS) Division at Ames Research Center for the production of the SPOC data products.

Some of the observations in this paper made use of the High-Resolution Imaging instrument `Alopeke and were obtained under Gemini LLP Proposal Number: GN/S-2019A-LP-105. `Alopeke was funded by the NASA Exoplanet Exploration Program and built at the NASA Ames Research Center by Steve B. Howell, Nic Scott, Elliott P. Horch, and Emmett Quigley. `Alopeke was mounted on the Gemini North telescope of the international Gemini Observatory, a program of NSF's OIR Lab, which is managed by the Association of Universities for Research in Astronomy (AURA) under a cooperative agreement with the National Science Foundation, on behalf of the Gemini partnership: the National Science Foundation (United States), National Research Council (Canada), Agencia Nacional de Investigación y Desarrollo (Chile), Ministerio de Ciencia, Tecnología e Innovación (Argentina), Ministério da Ciência, Tecnologia, Inovações e Comunicações (Brazil), and Korea Astronomy and Space Science Institute (Republic of Korea).

This publication makes use of data products from the Two Micron All Sky Survey, which is a joint project of the University of Massachusetts and the Infrared Processing and Analysis Center/California Institute of Technology, funded by the National Aeronautics and Space Administration and the National Science Foundation.

This work has made use of the following software: \texttt{radvel} \citep{2018PASP..130d4504F}, \texttt{astropy} \citep{2022ApJ...935..167A}, \texttt{matplotlib} \citep{2007CSE.....9...90H}, \texttt{numpy} \citep{2020Natur.585..357H}, \texttt{scipy} \citep{2020NaMet..17..261V}, and \texttt{lightkurve} \citep{2018ascl.soft12013L}.

\end{acknowledgements}
%
%

\bibliographystyle{aa} 
\bibliography{references} 

\begin{appendix}

\section{Frequency analysis}
\label{sec:frequency}

\subsection{Transit-like signals in the TESS photometry}

We verified the planet candidates reported by the TESS Science Office and searched for additional transit-like signals using the transit least squares algorithm \citep[\texttt{TLS;}][]{2019A&A...623A..39H}. For both targets, we first detrended the photometry using the biweight method (window length of 0.25 d) as implemented in \texttt{wotan} \citep{2019AJ....158..143H}, and combined the photometry from all sectors into a single light curve. Whenever a transit-like signal with a signal detection efficiency (SDE) greater than 10 was identified, we masked the photometry within 3 h of each mid-transit time and iteratively re-ran the \texttt{TLS} search.

For TOI-426, we detected TOI-426.01 ($P \approx 1.32$ d) with an SDE of 13.8, and no additional significant signals were found in subsequent iterations. For TOI-1839, we first identified TOI-1839.01 ($P \approx 1.42$ d) with an SDE of 44.7. In the second iteration, we detected an additional transit-like signal with a period of $P \approx 4.02$ d and an SDE of 45.9, which we denote as TOI-1839.02. This transit signal has also been independently identified and statistically validated by \citet{2026MNRAS.548ag512L}. No further signals were identified in the third iteration. %

\subsection{Stellar activity signals in the WASP photometry}
\label{sec:activity_wasp}

\begin{figure*}
    \centering
    \includegraphics[width=0.98\textwidth]{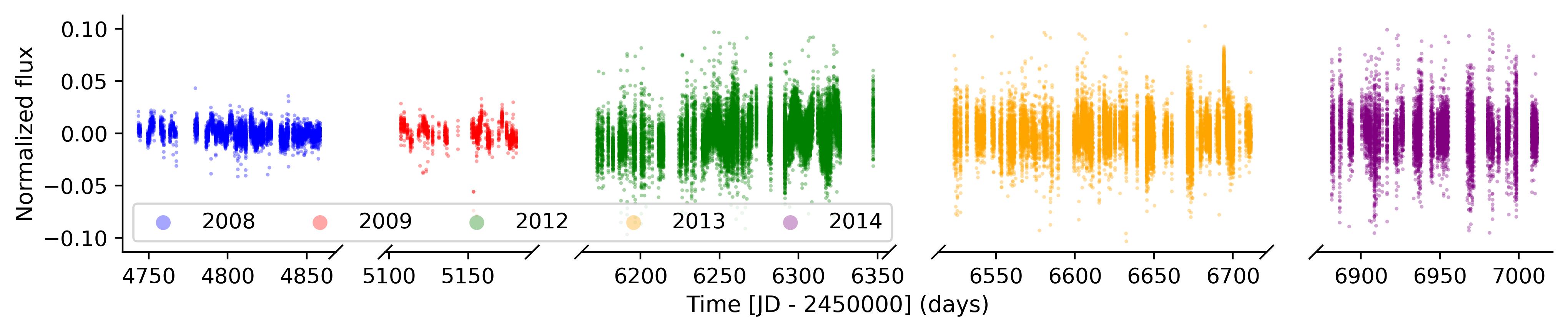}
    \includegraphics[width=0.98\textwidth]{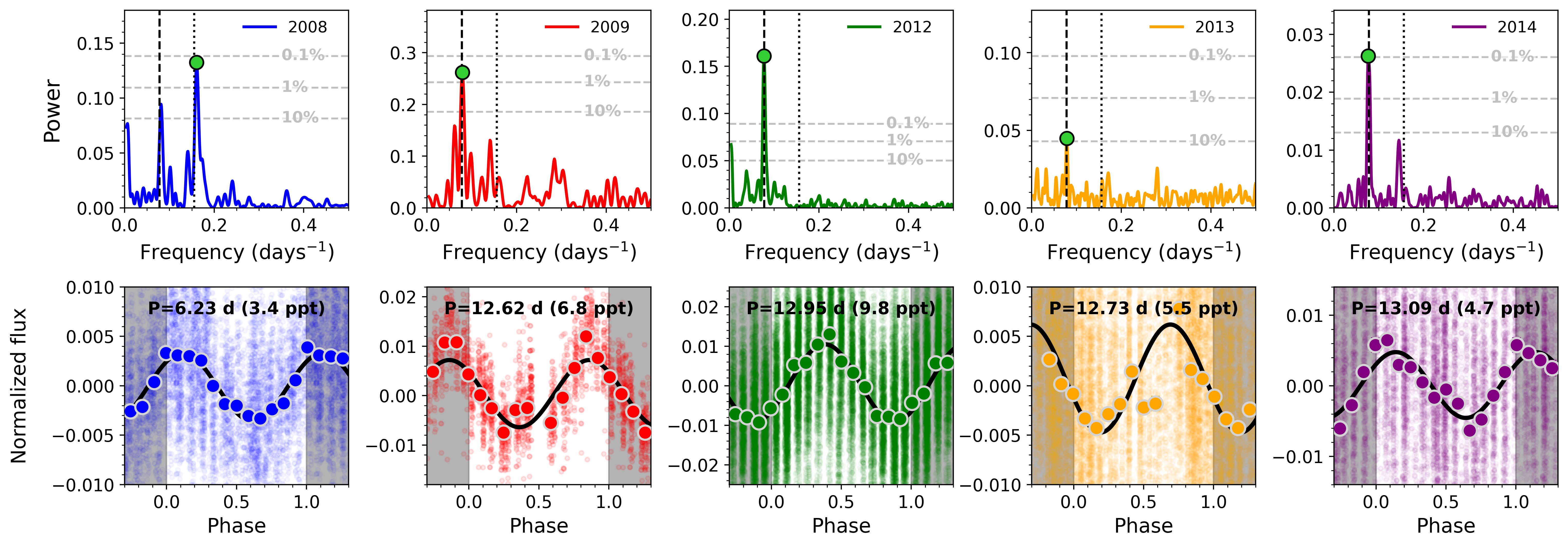}
    \caption{Top panel: WASP photometric time series of TOI-426. Middle panels: Year-by-year \texttt{GLS} periodograms, with circles marking the maximum-power frequencies (green for FAP $<10\%$, red otherwise). Vertical dashed and dotted lines indicate the mean frequency of $0.0779\,\mathrm{d}^{-1}$ (2009--2014) and its second harmonic, respectively, while horizontal lines mark the FAP levels. Bottom panels: Phase-folded light curves at the corresponding maximum-power periods, binned into 20 intervals (8\% in phase).}
    \label{fig:WASP_426}
\end{figure*}

\begin{figure*}
    \centering
    \includegraphics[width=0.98\textwidth]{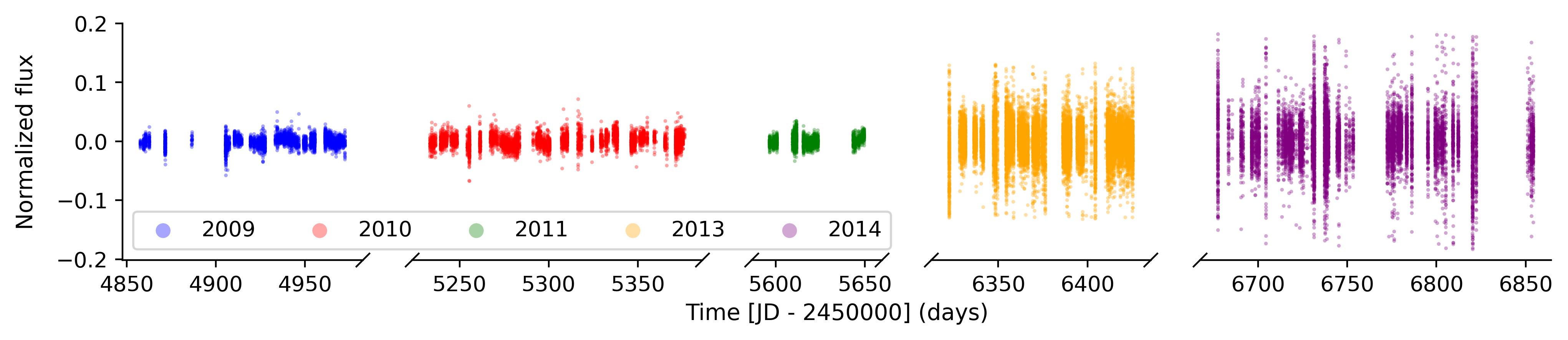}
    \includegraphics[width=0.98\textwidth]{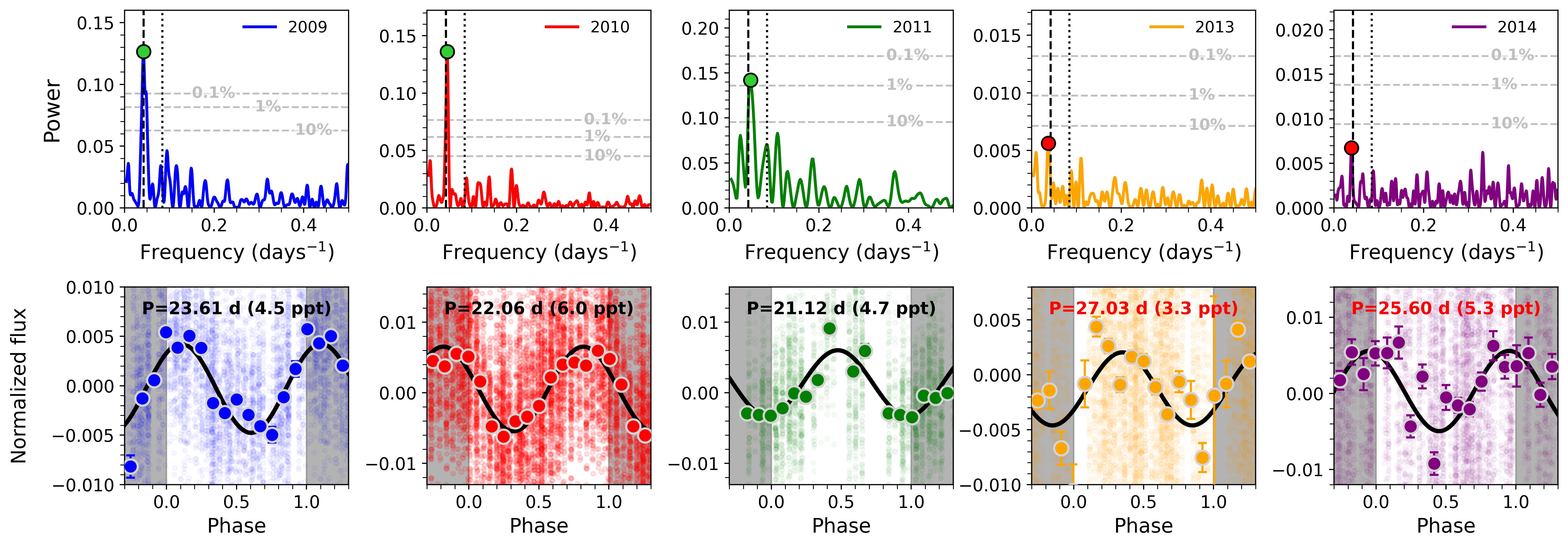}
    \caption{Top panel: WASP photometric time series of TOI-1839. Middle panels: Year-by-year \texttt{GLS} periodograms, with circles marking the maximum-power frequencies (green for FAP $<10\%$, red otherwise). Vertical dashed and dotted lines indicate the mean frequency of $0.0422\,\mathrm{d}^{-1}$ (2009--2014) and its second harmonic, respectively, while horizontal lines mark the FAP levels. Bottom panels: Phase-folded light curves at the corresponding maximum-power periods, binned into 20 intervals (8\% in phase).}
    \label{fig:WASP_1839}
\end{figure*}

We searched the WASP data for activity signals using the generalised Lomb-Scargle periodogram (\texttt{GLS}; \citealt{2009A&A...496..577Z}). In Figs.~\ref{fig:WASP_426} and~\ref{fig:WASP_1839}, we show the light curves, season-by-season periodograms, and phase-folded photometry. False-alarm probabilities (FAPs) were empirically derived using a bootstrap approach \citep[e.g.][]{Politis1994,2022A&A...658A.177H}.

The TOI-426 photometry exhibits a persistent modulation with a period of 12.6--13.1 d and an amplitude of 5--10 ppt. The signal is highly significant in 2012 and 2014 (FAP $<0.1\%$), significant in 2009 ($0.1\%<\mathrm{FAP}<1\%$), and tentative in 2013 (FAP $=9\%$).
In 2008, the second harmonic (6.2 d, 3.4 ppt, FAP $=0.15\%$) dominates the periodogram. We note that TOI-426 and its nearby companion TOI-949 lie within the photometric aperture, leading to dilution of the observed amplitudes. In particular, the WASP data alone do not allow us to determine which of the two stars is responsible for the detected modulation. TOI-1839 shows a stable modulation with a periodicity of 21.1--23.6 d and an amplitude of 5--6 ppt. The signal is highly significant in 2009 and 2010, and significant in 2011. In 2013 and 2014, the periodograms show no significant peaks (FAPs $>10\%$), which may be due to either the loss of photometric precision after the 2012 lens replacement or to a lower stellar activity level.

Overall, TOI-426 and TOI-1839 exhibit significant and long-lived photometric modulations at $\approx$ 13 and 22 d, respectively, which might trace their stellar rotation periods. In App.~\ref{sec:activity_HARPS}, we use the HARPS data to further assess the persistence of these signals and, in the case of TOI-426, to confirm their origin.

\subsection{Stellar activity and planetary signals in the HARPS data}
\label{sec:activity_HARPS}

\begin{figure*}
    \centering
    \includegraphics[width=\textwidth]{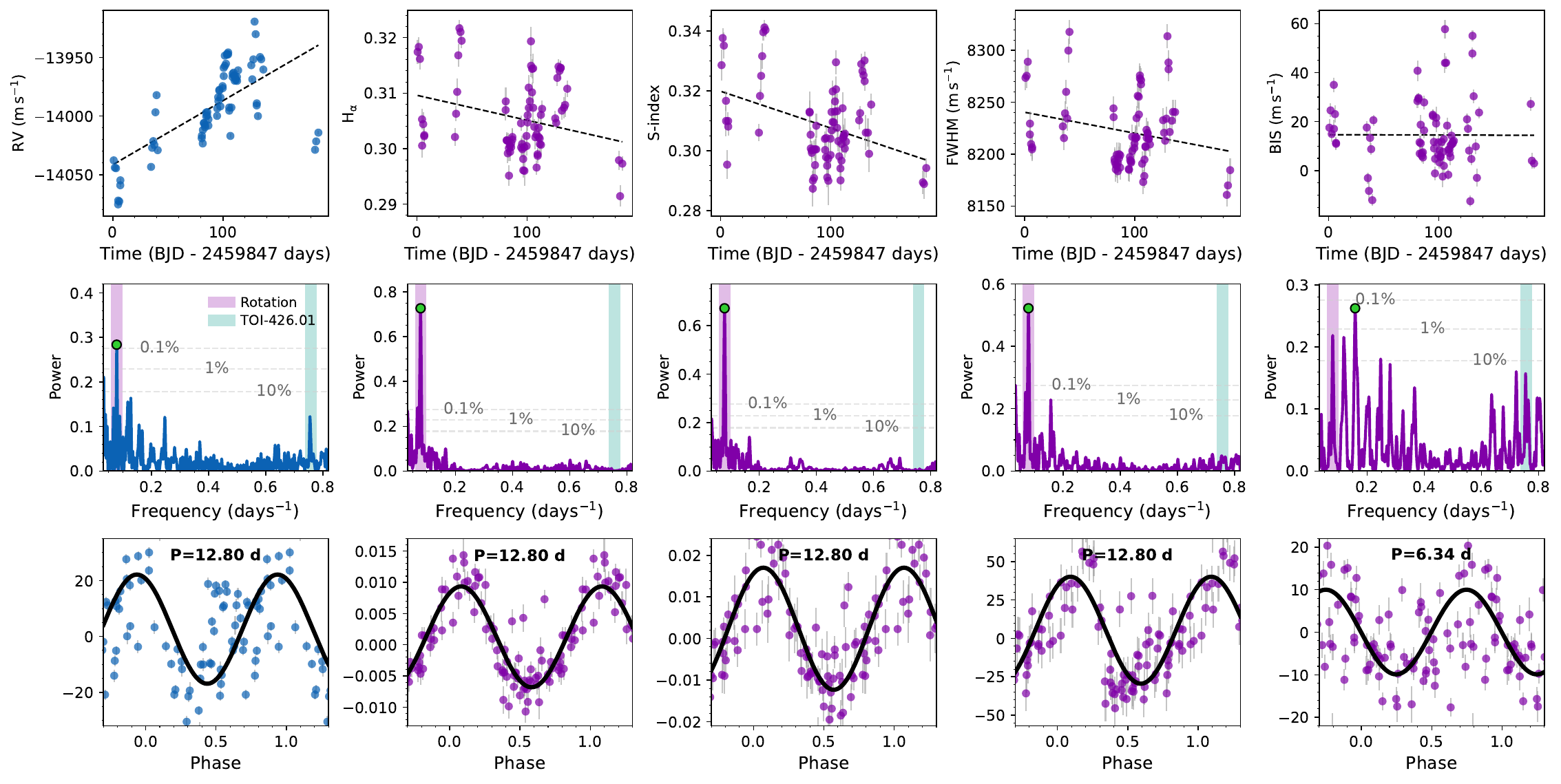}
    \caption{Upper panels: Time series of the HARPS RVs and activity indicators of TOI-426. The black dashed lines represent the linear trends fitted to the data. Middle panels: \texttt{GLS} periodograms of the de-trended time series. The green circles indicate the maximum power periods. The green vertical lines indicate the orbital period of TOI-426.01 ($P_{\rm orb}\approx1.32$ d). The magenta vertical lines indicate the 12.8-d persistent activity signal attributed to the stellar rotation period. The horizontal dashed lines correspond to the 10\%, 1\%, and 0.1\% FAP levels. Lower panels: De-trended time series phase-folded at the maximum power periods.}
    \label{fig:gls_to_HARPS_TOI-426}
\end{figure*}

\begin{figure*}
    \centering
    \includegraphics[width=\textwidth]{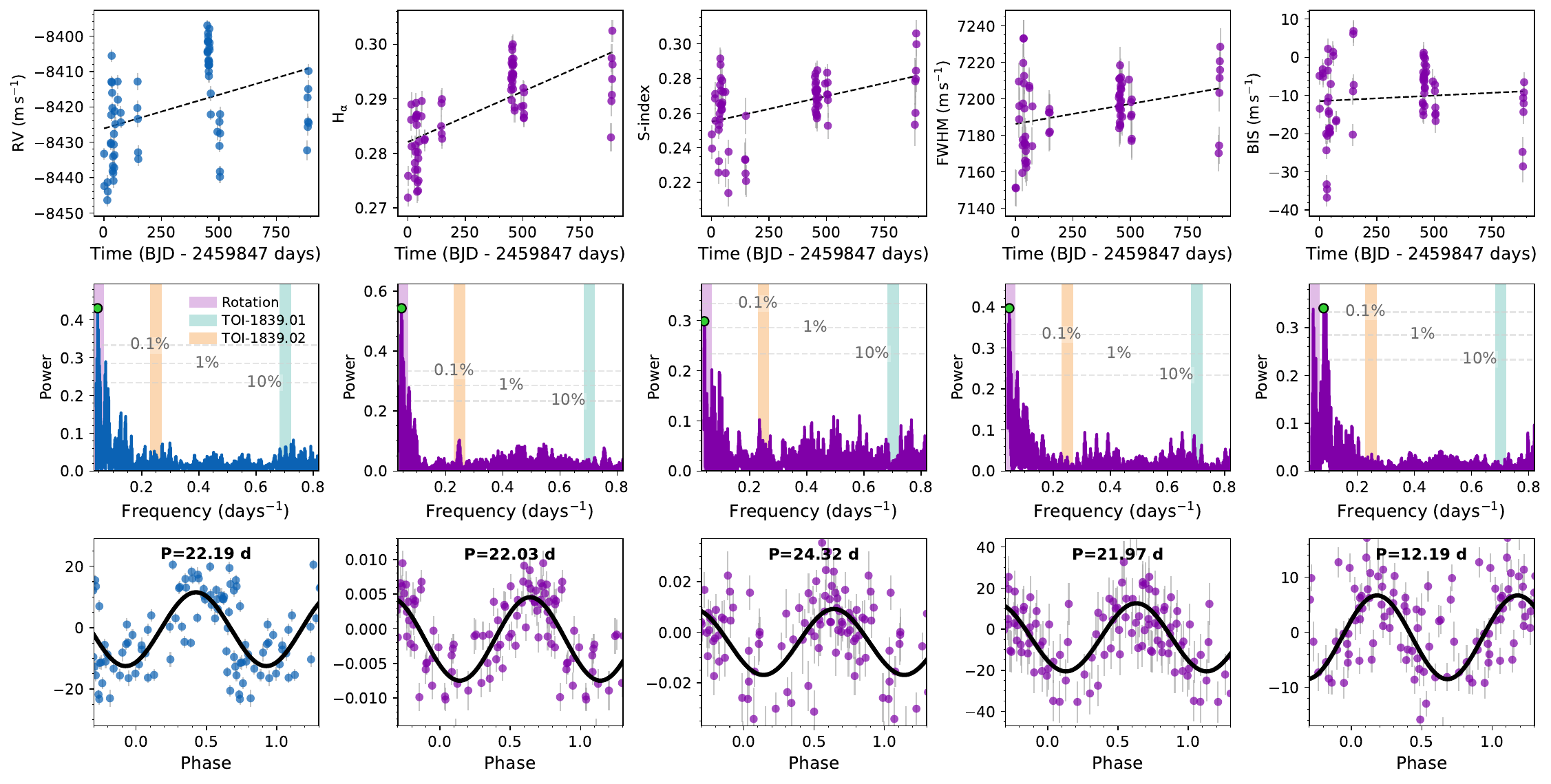}
    \caption{Upper panels: Time series of the HARPS RVs and activity indicators of TOI-1839. The black dashed lines represent the linear trends fitted to the data. Middle panels: \texttt{GLS} periodograms of the de-trended time series. The green circles indicate the maximum power periods. The green vertical lines mark the orbital period of TOI-1839.01 ($P_{\rm orb}\approx1.42$~d), while the orange vertical lines indicate the orbital period of TOI-1839.02 ($P_{\rm orb}\approx4.02$~d). The magenta vertical lines indicate the 21--23 d persistent activity signal attributed to the stellar rotation period. The horizontal dashed lines correspond to the 10\%, 1\%, and 0.1\% FAP levels. Lower panels: De-trended time series phase-folded at the maximum power periods.}
    \label{fig:gls_to_HARPS_TOI-1839}
\end{figure*}

We also used the \texttt{GLS} periodogram to search the HARPS data for planetary and activity signals. In Figs.~\ref{fig:gls_to_HARPS_TOI-426} and \ref{fig:gls_to_HARPS_TOI-1839}, we show the periodograms of the linearly de-trended RVs and four activity indicators (FWHM, BIS, H$\alpha$, and $S$ index) of TOI-426 and TOI-1839, respectively. The CCF contrast and the NaD doublet show a weaker response to stellar activity in our data set and were therefore not considered. The Ca index is largely redundant with the $S$ index, so it was also excluded from the analysis.

For TOI-426, the periodograms of the indicators H$\alpha$, $S$ index, and FWHM show maximum power periods of 12.8 d, which correspond to highly significant signals (FAPs $\ll 0.1\%$). The second harmonic arises as the main periodicity for the BIS indicator (6.3 d; FAP $= 0.2\%$). This behaviour is commonly observed for the BIS, which traces asymmetries in the CCF rather than bulk Doppler shifts. Stellar active regions can induce line-profile distortions that change sign twice per stellar rotation, often producing BIS variations dominated by the second harmonic of the rotation period ($P_{\rm rot}/2$; e.g. \citealt{2001A&A...379..279Q,2011A&A...528A...4B,2014ApJ...796..132D}). This recurrent and highly significant 12.8-d periodicity coincides with the signal detected in the WASP photometry (Sect.~\ref{sec:activity_wasp}), indicating the existence of an activity-related signal most likely caused by the rotation of TOI-426. The RV periodogram shows the same 12.8-d signal with a significant FAP of 0.07\%, indicating that an important component of the RV variations has a stellar origin. We also note that the RV periodogram of TOI-426 shows a tentative peak (FAP $> 10\%$) at the 1.3-d periodicity of the TESS candidate TOI-426.01 with no counterpart in the indicators, suggesting that its signature is embedded in the RVs.

For TOI-1839, the periodograms of H$\alpha$, FWHM, and $S$ index show maximum power periods between 22 and 24 d with FAPs between 0.6\% ($S$ index) and $3 \times 10^{-7}\%$ (H$\alpha$). The second harmonic also arises as the main periodicity for the BIS indicator (12.2 d; FAP $= 0.07\%$). This 22-d periodicity coincides with the signal detected in the WASP photometry (Sect.~\ref{sec:activity_wasp}) and the main periodicity in the RV time series (FAP of $5 \times 10^{-4}\%$), indicating that stellar rotation also plays an important role in shaping the RV variability of TOI-1839.

As shown in Fig.~\ref{fig:gls_to_HARPS_TOI-426}, the RVs of TOI-426 show a strong upward trend, whereas the FWHMs exhibit a mild downward trend. Since long-term variations caused by stellar magnetic cycles often produce trends of the same sign in the RVs and FWHMs \citep[e.g.][]{2011arXiv1107.5325L}, the opposite signs observed in TOI-426 suggest that the RV trend might instead be produced by a long-period companion. Along these lines, we note that the three most recent RV measurements deviate significantly from the linear trend, while these measurements broadly follow the downward tendency in all the activity indicators, further suggesting that the long-term behaviours of the RVs and activity indicators are uncorrelated. For TOI-1839, we observe mild trends in the RVs and FWHMs with the same sign (Fig.~\ref{fig:gls_to_HARPS_TOI-1839}), preventing us from drawing conclusions about their possible origin. %

\section{Interior structure and atmospheric escape of TOI-426\,b, TOI-1839\,b, and TOI-1839\,c}
\label{sec:interior_escape}

\subsection{Interior modelling}
\label{sec:interior-modeling}

We computed the bulk compositions of TOI-426\,b, TOI-1839\,b, and TOI-1839\,c considering three classes of interior structures: (a) gas dwarfs with $1\times$ solar metallicity envelopes, (b) volatile-enriched envelopes with $50\times$ solar metallicity, and (c) envelopes composed of pure \ce{H2O} in a steam or supercritical state. For the hydrogen-dominated cases, we used the grid of interior structure models from \cite{2025ApJ...989...28T}. For steam worlds, we adopted the models from \cite{2025ApJ...988..186A}. The original grid applies to planets with equilibrium temperatures between 500 and 700~K, while an extended grid covering 400--1500~K is available at \cite{aguichine_2025_17811236}. Both model grids were smoothly connected to the Earth-like radius from \cite{2016ApJ...819..127Z} in the limit of zero envelope mass fraction, following \cite{2021AJ....161...70P}. Both model families describe the thermal evolution of planets composed of an Earth-like core surrounded by a volatile envelope, which experience a hot start during formation and subsequently cool by radiating energy through the atmosphere. As the interior loses heat, the planet contracts and its radius decreases with time. These are fully coupled atmosphere–interior models, as they simultaneously track the thermal energy stored in the interior and the rate at which this energy is transported through and lost from the atmosphere.

The planetary compositions were inferred within a Bayesian framework using MCMC sampling. We adopted Gaussian priors on the planet mass and equilibrium temperature, as reported in Table~\ref{tab:toi426_toi1839_priors_posteriors}, and a uniform prior on the system age within the $1\sigma$ interval derived from gyrochronology (Sect.~\ref{subsec:age}, Table~\ref{tab:stellar_properties}). The observed planetary radius was incorporated as a Gaussian term in the likelihood function $\mathcal{L}$. For the steam-world models \citep{2025ApJ...988..186A}, we assumed a uniform prior on the bulk water mass fraction (WMF) between 0 and 100\%. For the gas-dwarf models \citep{2025ApJ...989...28T}, we sampled the envelope mass fraction using the transformed variable $y = \ln f_{\mathrm{env}}$ and included the corresponding Jacobian term, $\ln \left| \mathrm{d}f_{\mathrm{env}} / \mathrm{d}y \right| = y$, in the log-likelihood $\ln \mathcal{L}$, ensuring an effectively uniform prior in $f_{\mathrm{env}}$. The MCMC sampling was performed with \texttt{emcee} \citep{2013PASP..125..306F}, yielding posterior distributions for the WMF and $f_{\mathrm{env}}$. The inferred envelope mass fractions and total envelope masses for TOI-426\,b, TOI-1839\,b, and TOI-1839\,c are summarised in Table~\ref{tab:volatile-content-modelling-results} and are consistent with their positions in the mass--radius diagram (Fig.~\ref{fig:mr-plotter}).

\subsection{Atmospheric escape}
\label{sec:escape-modeling}

We computed and integrated the atmospheric mass-loss rates of TOI-426\,b, TOI-1839\,b, and TOI-1839\,c following the methodology presented by \citet{2021ApJ...914...84A}. Over the system lifetime, atmospheric escape is primarily regulated by the evolution of the stellar XUV irradiation that drives photoevaporation. We therefore integrated the mass-loss rates assuming fixed planetary mass, radius, and equilibrium temperature, while allowing the stellar XUV luminosity to evolve with time. To account for model-dependent uncertainties, we computed mass-loss rates using different prescriptions: the analytical fit from \cite{2017ApJ...847...29O} (OW17), the analytical fit from \cite{2016A&A...586A..75S} (S16), and the grid-based calculations from \cite{2018A&A...619A.151K,2021RNAAS...5...74K} (K21). Uncertainties were estimated through Monte Carlo sampling with $10^4$ realisations, adopting the same priors on planetary mass, radius, equilibrium temperature, and age as in the interior modelling. The results are summarised in Table~\ref{tab:volatile-content-modelling-results}. The OW17 prescription, which represents an approximate fit to numerical simulations, yields systematically lower escape efficiencies, whereas the S16 and K21 prescriptions provide mutually consistent results.

In the gas-dwarf scenario (i.e. \ce{H2}/He-dominated envelopes), the total mass lost over the system lifetime exceeds the present-day envelope mass ($\Delta M \gg M_{\mathrm{env}}$), implying that primordial envelopes would be efficiently removed. Using the S16 and K21 prescriptions, the corresponding mass-loss timescales are of order $\sim1$--100\,Myr. The OW17 prescription yields longer timescales ($\sim0.1$--2\,Gyr), although these should be interpreted with caution given the limitations of the model.

In the steam-world scenario, no dedicated grids of \ce{H2O} escape rates are currently available. We therefore estimated present-day escape rates using \texttt{Wind-AE} \citep{2025ApJ...995..198B}. This model solves the atmospheric structure above the optical transit radius, and the mass-loss rate corresponds to the mass outflow at the sonic point. In this part of the atmosphere, water is fully dissociated, so we assumed the atmosphere to be composed of H and O in the proportions corresponding to \ce{H2O}. The detailed methodology is described by Tao et al. (in prep). The resulting values correspond to mass-loss timescales of $\sim$100--1000\,Gyr, indicating that \ce{H2O}-dominated envelopes can be retained over the system lifetimes. Assuming that instantaneous and integrated mass loss scale similarly across compositions, we estimated that $\sim 0.5\,\mathrm{M_\oplus}$ of \ce{H2O} may have been removed from TOI-426\,b and TOI-1839\,b, and $\sim 0.1\,\mathrm{M_\oplus}$ from TOI-1839\,c over their lifetimes. These values are small compared to the inferred volatile contents, supporting the stability of \ce{H2O}-dominated envelopes in these planets.

\begin{table}[]
\centering
\renewcommand{\arraystretch}{1.75}
\setlength{\tabcolsep}{8.0pt}
\caption{Compositions of TOI-426\,b, TOI-1839\,b, and TOI-1839\,c inferred from interior structure modelling (Sect.~\ref{sec:interior-modeling}), and corresponding atmospheric escape properties derived from the escape modelling (Sect.~\ref{sec:escape-modeling}).}
\label{tab:volatile-content-modelling-results}
\begin{tabular}{lccc} 
\hline \hline
  & TOI-426\,b & TOI-1839\,b & TOI-1839\,c \\ \hline
\multicolumn{4}{l}{Mass fraction of the envelope (\%)} \\ \hline
$f_{\mathrm{env,1}}$      & $0.033^{+0.012}_{-0.014}$          &   $0.049^{+0.017}_{-0.017}$   &    $0.14^{+0.13}_{-0.06}$       \\
$f_{\mathrm{env,L}}$      & $0.059^{+0.028}_{-0.027}$           & $0.18^{+0.21}_{-0.10}$          & $0.57^{+0.51}_{-0.33}$          \\
$f_{\mathrm{env,H_2O}}$    & $22^{+16}_{-12}$          & $ 33^{+17}_{-14}$           & $ 51^{+21}_{-19}$           \\  \hline
\multicolumn{4}{l}{Mass of the envelope ($\mathrm{M_\oplus}$)} \\ \hline
$M_{\mathrm{env,1}}$      & $0.0022^{+0.0010}_{-0.0009}$          &   $0.0035^{+0.0013}_{-0.0012}$   &    $0.0084^{+0.0081}_{-0.0040}$       \\
$M_{\mathrm{env,L}}$      & $0.0040^{+0.0019}_{-0.0016}$           & $0.013^{+0.014}_{-0.007}$          & $0.033^{+0.031}_{-0.019}$          \\
$M_{\mathrm{env,H_2O}}$    & $1.4^{+0.6}_{-0.6}$          & $ 2.3^{+1.1}_{-1.0}$           & $ 3^{+1.1}_{-1.0}$           \\  \hline
\multicolumn{4}{l}{Present-day mass-loss rate of \ce{H2}/He ($\mathrm{M_\oplus\,Gyr^{-1}}$)} \\ \hline
$\dot{m}_{\mathrm{OW17}}$ & $0.028^{+0.038}_{-0.015}$           & $0.013^{+0.008}_{-0.005}$         & $0.005^{+0.004}_{-0.002}$         \\
$\dot{m}_{\mathrm{S16}}$ & $1.4^{+1.5}_{-0.7}$     & $1.1^{+0.5}_{-0.3}$          & $0.32^{+0.20}_{-0.12}$          \\
$\dot{m}_{\mathrm{K21}}$ & $1.7^{+8.4}_{-1.2}$          & $4.2^{+3.3}_{-1.7}$           & $1.5^{+1.1}_{-0.5}$           \\ \hline 
\multicolumn{4}{l}{Present-day mass-loss rate of \ce{H2O} ($\mathrm{M_\oplus\,Gyr^{-1}}$)} \\ \hline
$\dot{m}_{\mathrm{Wind-AE}}$ & $0.01157$      & $0.00821$       & $0.00295$       \\ \hline
\multicolumn{4}{l}{Total mass of \ce{H2}/He lost since formation ($\mathrm{M_\oplus}$)} \\ \hline
$\Delta M_{\mathrm{OW17}}$ & $0.69^{+0.68}_{-0.29} $     & $0.74^{+0.32}_{-0.23} $   & $0.29^{+0.19}_{-0.11} $   \\
$\Delta M_{\mathrm{S16}}$ & $11^{+9}_{-4} $       & $11^{+5}_{-3} $      & $3.3^{+2.0}_{-1.2} $      \\
$\Delta M_{\mathrm{K21}}$    & $4.3^{+10.8}_{-2.8} $           & $ 4.2^{+3.3}_{-1.7} $ & $1.5^{+1.1}_{-0.5} $ \\
\hline
\end{tabular}
\end{table}

\section{Additional figures}

\begin{figure*}
    \centering
    \includegraphics[width=0.97\textwidth]{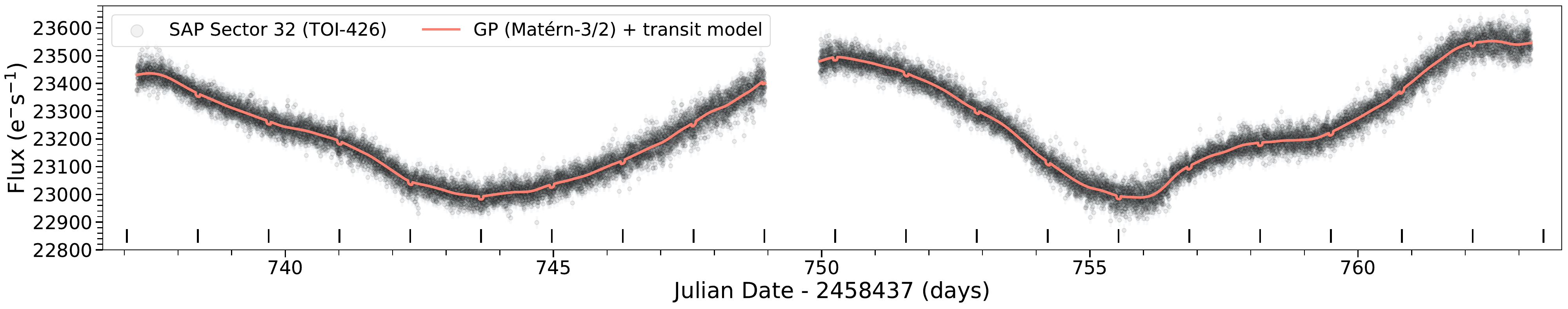}
    
    \vspace{0.15cm}
    
    \includegraphics[width=0.97\textwidth]{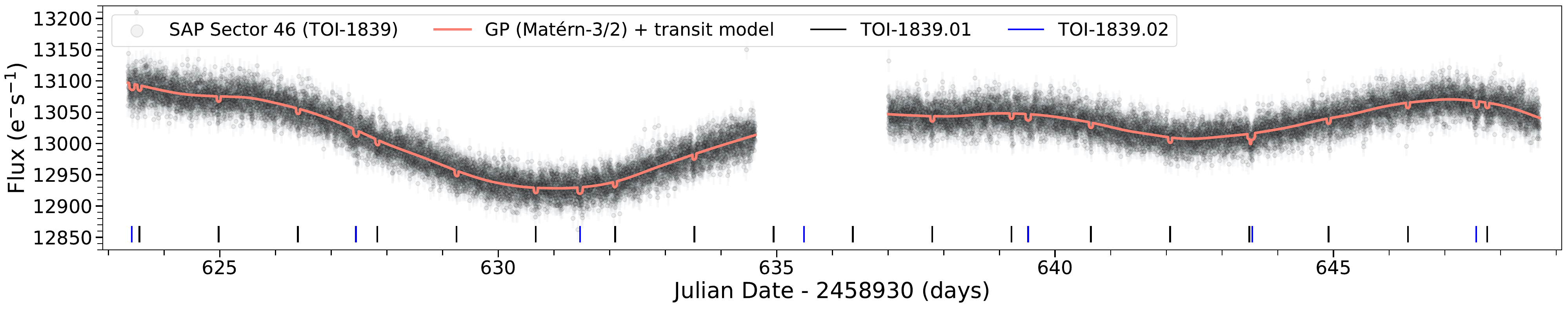}
    
    \caption{Example TESS sectors for TOI-426 (top; Sector 32) and TOI-1839 (bottom; Sector 46) together with the corresponding median posterior models (transit + GP) inferred in Sect.~\ref{subsec:analysis_TESS}. The vertical black lines indicate the derived transit ephemerides.}
    
    \label{fig:TESS_example_sectors}
\end{figure*}

\begin{figure*}
    \centering
    \includegraphics[width=0.97\textwidth]{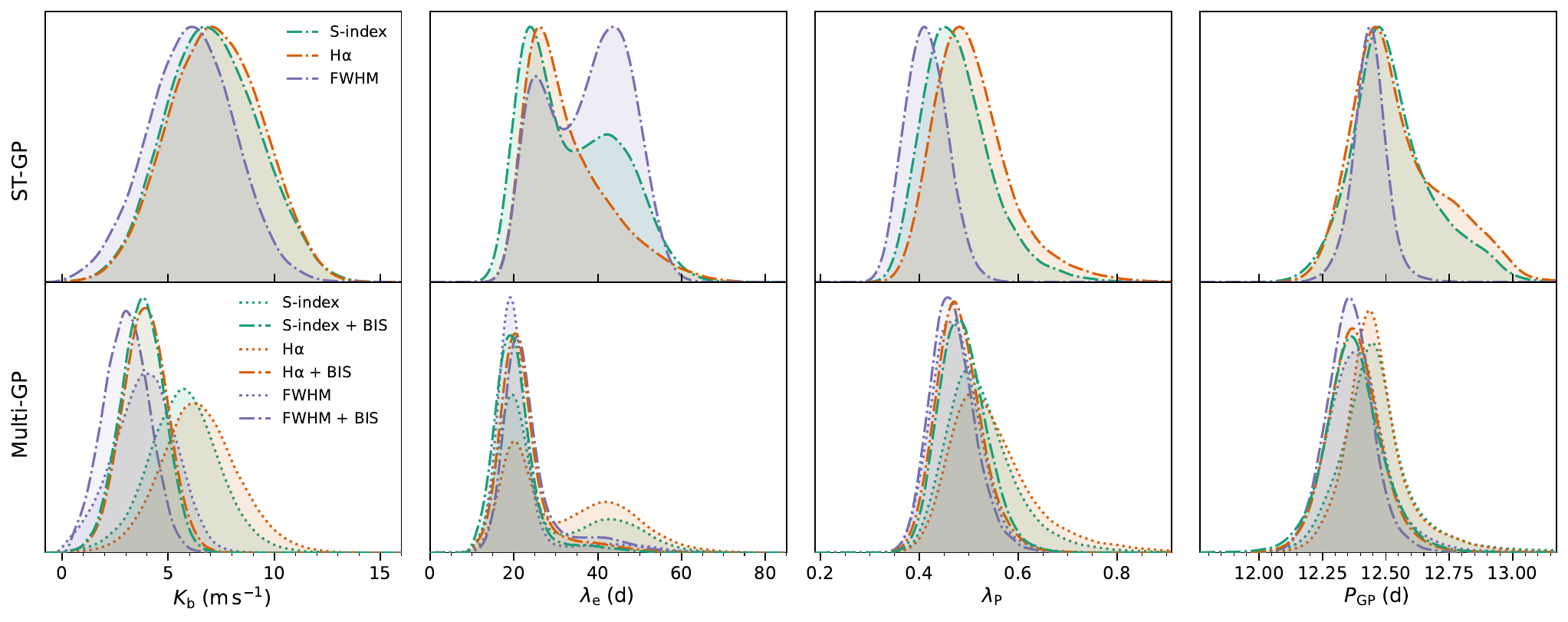}
    \includegraphics[width=0.972\textwidth]{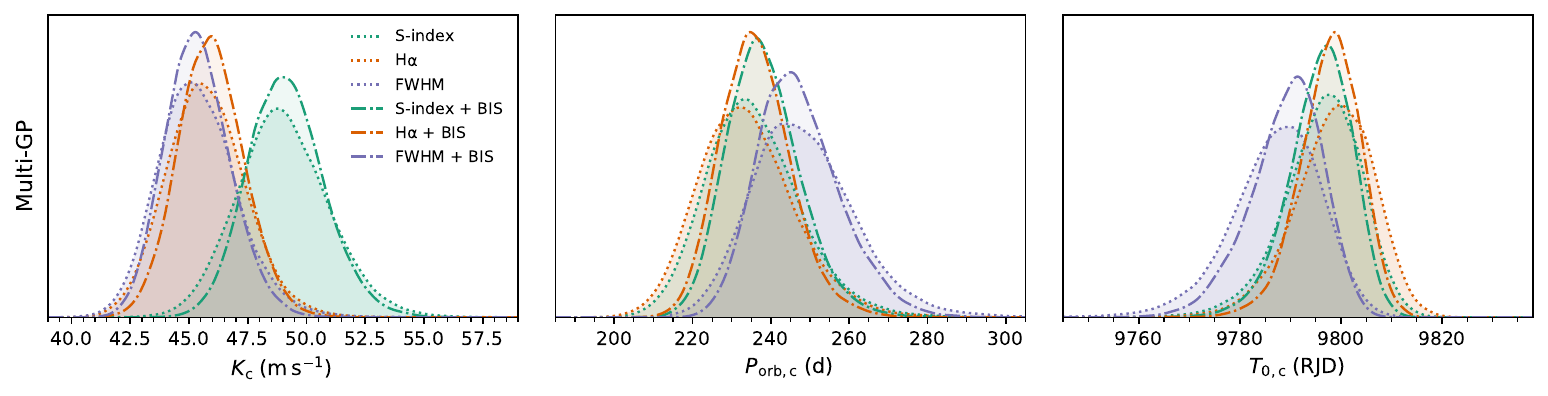}
    \caption{Posterior distributions of the Keplerian semi-amplitudes in the TOI-426 system ($K_{\rm b}$ and $K_{\rm c}$) and the quasi-periodic GP hyperparameters ($\lambda_{\rm e}$, $\lambda_{\rm P}$, and $P_{\rm GP}$) for the preferred ST-GP and multi-GP models. The preferred ST-GP model includes only TOI-426\,b on a circular orbit, whereas the preferred multi-GP model includes both TOI-426\,b and TOI-426\,c on circular orbits.}
    \label{fig:posterior_426}
\end{figure*}

\begin{figure*}
    \centering
    \includegraphics[width=\textwidth]{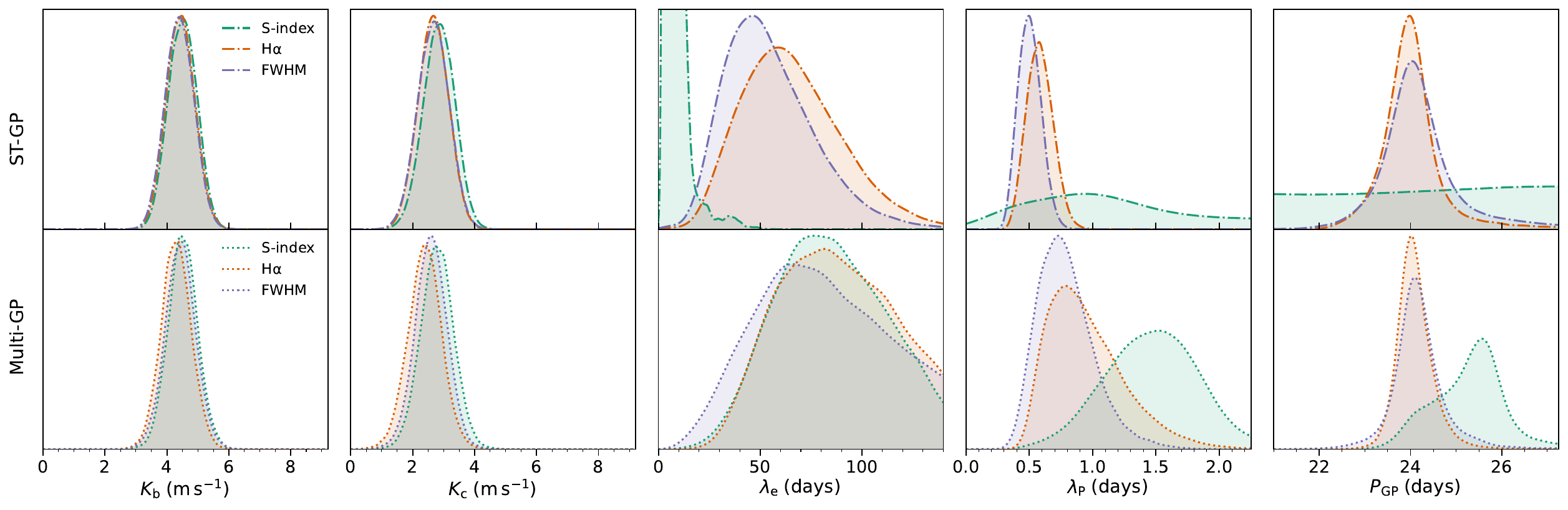}
    \caption{Posterior distributions of the Keplerian semi-amplitudes of TOI-1839\,b and TOI-1839\,c ($K_{\rm b}$ and $K_{\rm c}$), together with the hyperparameters of the quasi-periodic Gaussian process ($\lambda_{\rm e}$, $\lambda_{\rm P}$, and $P_{\rm GP}$), for the preferred two-planet circular models obtained from the ST-GP and 2D-GP analyses.}
    \label{fig:posterior_1839}
\end{figure*}

\begin{figure*}
    \centering
    \includegraphics[width=0.99\textwidth]{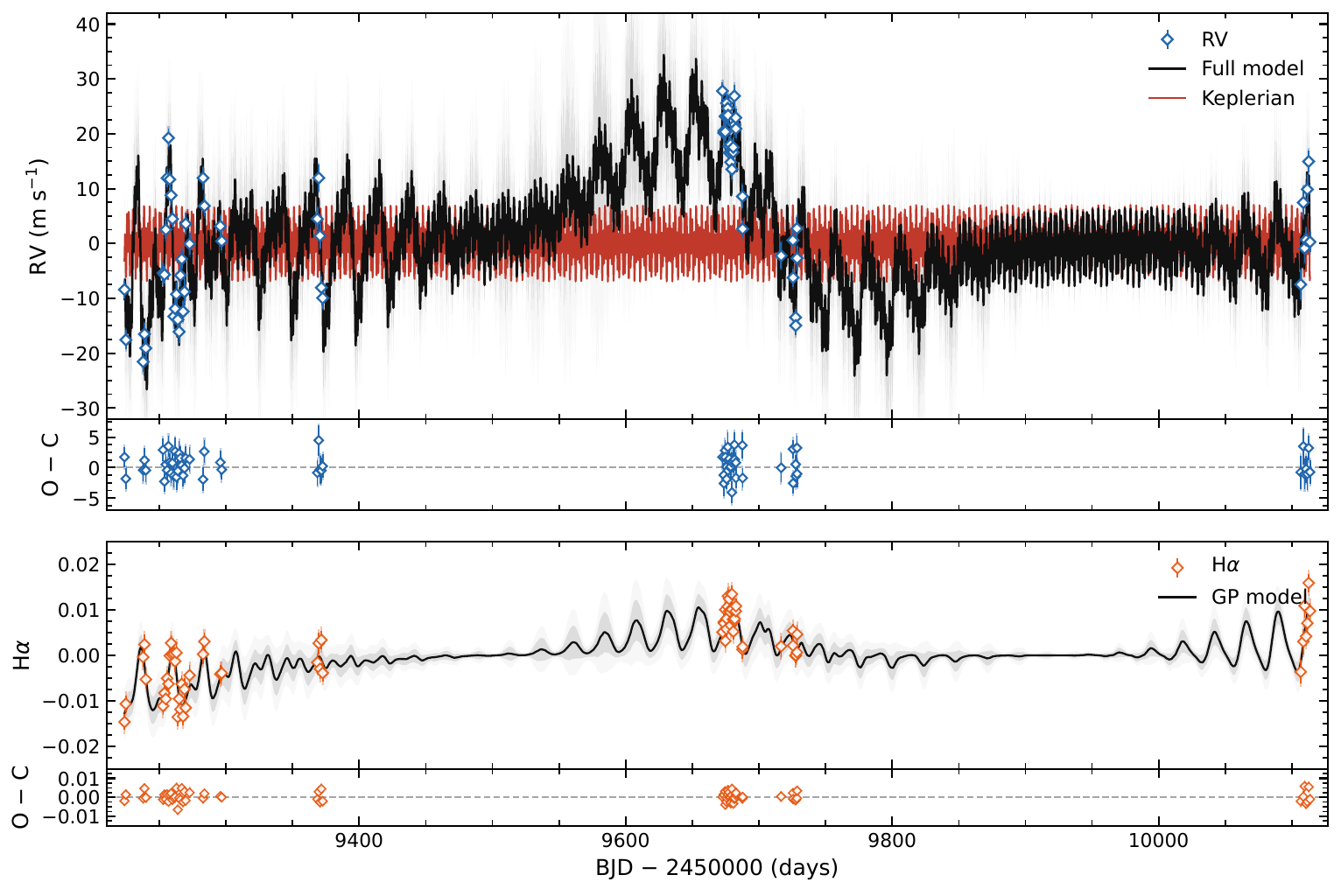}
    \caption{Full RV model for TOI-1839. Top panel: HARPS RVs, with the black and red lines showing the median posterior full and Keplerian models, respectively. The Keplerian model corresponds to the sum of the TOI-1839\,b and TOI-1839\,c signals. Bottom panel: HARPS H$\alpha$ measurements, with the black line showing the median posterior GP model. In both panels, the time series were corrected for the inferred instrumental offsets, and the dark and light grey bands denote the $1\sigma$ and $2\sigma$ confidence intervals.}
    \label{fig:1839_RV_full}
\end{figure*}

\newpage

\section{Additional tables}

\onecolumn

\begin{table}[h]
\centering
\renewcommand{\arraystretch}{1.15}
\setlength{\tabcolsep}{14pt}
\caption{Chemical abundances of TOI-426 and TOI-1839 derived in Sect.~\ref{subsec:atmospheric_parameters}.}
\begin{tabular}{lll}
\hline \hline
Parameter & TOI-426 & TOI-1839 \\
\hline
$[\mathrm{C/H}]$ (dex)  & $0.19 \pm 0.07$ & $0.05 \pm 0.09$ \\
$[\mathrm{O/H}]$ (dex)  & $0.17 \pm 0.06$ & $0.02 \pm 0.08$ \\
$[\mathrm{Mg/H}]$ (dex) & $0.11 \pm 0.04$ & $0.10 \pm 0.04$ \\
$[\mathrm{Si/H}]$ (dex) & $0.10 \pm 0.04$ & $0.07 \pm 0.02$ \\
$[\mathrm{Ti/H}]$ (dex) & $0.15 \pm 0.03$ & $0.12 \pm 0.04$ \\
$[\mathrm{Ni/H}]$ (dex) & $0.11 \pm 0.02$ & $0.08 \pm 0.03$ \\
$[\mathrm{Cu/H}]$ (dex) & $0.18 \pm 0.06$ & $0.06 \pm 0.07$ \\
$[\mathrm{Zn/H}]$ (dex) & $0.14 \pm 0.05$ & $0.09 \pm 0.06$ \\
$[\mathrm{Sr/H}]$ (dex) & $0.21 \pm 0.08$ & $0.04 \pm 0.09$ \\
$[\mathrm{Y/H}]$ (dex)  & $0.25 \pm 0.06$ & $0.04 \pm 0.06$ \\
$[\mathrm{Zr/H}]$ (dex) & $0.23 \pm 0.07$ & $0.13 \pm 0.07$ \\
$[\mathrm{Ba/H}]$ (dex) & $0.28 \pm 0.03$ & $0.04 \pm 0.03$ \\
$[\mathrm{Ce/H}]$ (dex) & $0.22 \pm 0.03$ & $0.11 \pm 0.06$ \\
$[\mathrm{Nd/H}]$ (dex) & $0.19 \pm 0.03$ & $0.07 \pm 0.03$ \\
$A(\mathrm{Li})$ (dex)  & $2.00 \pm 0.04$ & $<0.3$ \\
\hline
\end{tabular}
\label{tab:chemical_abundances}
\end{table}

\begin{table}[h]
\centering
\renewcommand{\arraystretch}{1.15}
\setlength{\tabcolsep}{19.6pt}
\caption{Matérn-3/2 GP hyperparameters inferred from the TESS photometry analysis described in Sect.~\ref{subsec:analysis_TESS}.}
\begin{tabular}{lllll}
\hline\hline
& \multicolumn{2}{c}{TOI-426} & \multicolumn{2}{c}{TOI-1839} \\
\hline
Parameter & Priors & Posteriors & Priors & Posteriors \\
\hline
\multicolumn{5}{l}{Matérn-3/2 GP hyperparameters} \\
\hline
$\eta_{\rm \sigma_{S_X}}$ ($\mathrm{e^{-}\,s^{-1}}$)
& $\mathcal{U}(0,1000)$
& $438^{+299}_{-192}$
& $\mathcal{U}(0,1000)$
& $43.4^{+9.2}_{-6.5}$ \\

$\eta_{\rm \sigma_{S_Y}}$ ($\mathrm{e^{-}\,s^{-1}}$)
& $\mathcal{U}(0,1000)$
& $466^{+288}_{-185}$
& $\mathcal{U}(0,1000)$
& $498^{+324}_{-266}$ \\

$\eta_{\rm \sigma_{S_Z}}$ ($\mathrm{e^{-}\,s^{-1}}$)
& $\mathcal{U}(0,1000)$
& $292^{+81}_{-55}$
& $\mathcal{U}(0,1000)$
& $72^{+181}_{-33}$ \\

$\eta_{\rm \rho_{S_X}}$ (d)
& $\mathcal{U}(0,100)$
& $6.94^{+3.06}_{-2.35}$
& $\mathcal{U}(0,100)$
& $0.88^{+0.17}_{-0.13}$ \\

$\eta_{\rm \rho_{S_Y}}$ (d)
& $\mathcal{U}(0,100)$
& $9.00^{+3.55}_{-2.71}$
& $\mathcal{U}(0,100)$
& $31.3^{+13.1}_{-12.8}$ \\

$\eta_{\rm \rho_{S_Z}}$ (d)
& $\mathcal{U}(0,100)$
& $3.36^{+0.65}_{-0.48}$
& $\mathcal{U}(0,100)$
& $5.11^{+7.06}_{-1.90}$ \\
\hline
\end{tabular}
\tablefoot{For TOI-426, $S_X$, $S_Y$, and $S_Z$ denote TESS Sectors 5, 32, and 98, respectively. For TOI-1839, $S_X$, $S_Y$, and $S_Z$ denote TESS Sectors 23, 46, and 50, respectively.}
\label{tab:TESS_GP_hyperparams}
\end{table}

\begin{table*}[h]
\centering
\renewcommand{\arraystretch}{1.15}
\caption{Ranking of models by increasing \mgic\ for TOI-426.}
\label{tab:mgic_toi426}
\begin{tabular}{lrrrr}
\hline \hline
Model & $\mathcal{L}^\prime_{\rm rv}$ & \ksmooth\ & \kpar\ & \mgic\ \\
\hline
3D-GP - RV + FWHM + BIS - extra circular Keplerian & 298 & 36 & 20 & -484 \\
3D-GP - RV + $H_{\alpha}$ + BIS - extra circular Keplerian & 300 & 38 & 20 & -483 \\
3D-GP - RV + $S$ index + BIS - extra circular Keplerian & 301 & 39 & 20 & -483 \\
3D-GP - RV + FWHM + BIS - eccentric Keplerian & 298 & 35 & 22 & -481 \\
3D-GP - RV + $S$ index + BIS - eccentric Keplerian & 302 & 39 & 22 & -481 \\
3D-GP - RV + $H_{\alpha}$ + BIS - eccentric Keplerian & 298 & 36 & 22 & -480 \\
2D-GP - RV + FWHM - extra circular Keplerian & 290 & 42 & 16 & -465 \\
2D-GP - RV + $H_{\alpha}$ - extra circular Keplerian & 266 & 34 & 16 & -432 \\
2D-GP - RV + $S$ index - extra circular Keplerian & 272 & 40 & 16 & -432 \\
2D-GP - RV + FWHM - quadratic trend & 225 & 6 & 15 & -407 \\
1D-GP - RV - extra circular Keplerian & 257 & 45 & 12 & -398 \\
1D-GP - RV - quadratic trend & 254 & 45 & 11 & -395 \\
1D-GP - RV - no trends & 247 & 42 & 9 & -391 \\
2D-GP - RV + FWHM - no trends & 157 & 1 & 13 & -287 \\
\hline
\end{tabular}
\end{table*}

\begin{table}[htbp]
\centering
\renewcommand{\arraystretch}{1.15}
\caption{Ranking of models by increasing \mgic\ for TOI-1839.}
\label{tab:mgic_toi1839}
\begin{tabular}{lrrrr}
\hline \hline
Model & $\mathcal{L}^\prime_{\rm rv}$ & \ksmooth\ & \kpar\ & \mgic\ \\
\hline
2D-GP - RV + $H_{\alpha}$ & 276 & 25 & 16 & -470 \\
2D-GP - RV + FWHM & 275 & 25 & 16 & -468 \\
2D-GP - RV + $S$ index & 269 & 25 & 16 & -458 \\
1D-GP - RV & 269 & 34 & 9 & -452 \\
\hline
\end{tabular}
\end{table}

\renewcommand{\arraystretch}{1.25}
\begin{longtable}{lllll}
\caption{Priors and posterior distributions for TOI-426 and TOI-1839.}
\label{tab:toi426_toi1839_priors_posteriors} \\

\hline\hline
 & \multicolumn{2}{c}{TOI-426} & \multicolumn{2}{c}{TOI-1839} \\
\hline
Parameter & Prior & Posterior & Prior & Posterior \\
\hline
\endfirsthead

\multicolumn{5}{c}{{\tablename\ \thetable{} -- continued}} \\
\hline\hline
 & \multicolumn{2}{c}{TOI-426} & \multicolumn{2}{c}{TOI-1839} \\
\hline
Parameter & Prior & Posterior & Prior & Posterior \\
\hline
\endhead

\hline
\multicolumn{5}{r}{{Continued on next page}} \\
\endfoot

\hline
\endlastfoot

\multicolumn{5}{l}{Planet b $-$ sampled parameters} \\
\hline
$T_{0,\mathrm{b}}$ (RJD) &
$\mathcal{U}(8438.52,8438.54)$ &
$8438.52836^{+0.00084}_{-0.00077}$ &
$\mathcal{U}(9690.2,9690.3)$ &
$9690.2390^{+0.0010}_{-0.0012}$ \\

$P_{\mathrm{b}}$ (d) &
$\mathcal{U}(1.320,1.321)$ &
$1.32050258^{+0.00000075}_{-0.00000084}$ &
$\mathcal{U}(1.41,1.43)$ &
$1.4237828 \pm 0.0000045$ \\

$R_{\mathrm{p,b}}/R_{\star}$ &
$\mathcal{U}(0.0,0.1)$ &
$0.02028 \pm 0.00053$ &
$\mathcal{U}(0,0.1)$ &
$0.02337 \pm 0.00074$ \\

$K_{\mathrm{b}}$ (m\,s$^{-1}$) &
$\mathcal{U}(0,100)$ &
$3.9^{+1.0}_{-1.1}$ &
$\mathcal{U}(0,100)$ &
$4.44^{+0.46}_{-0.45}$ \\

$b_{\mathrm{b}}$ &
$\mathcal{U}(0,1)$ &
$0.269^{+0.093}_{-0.133}$ &
$\mathcal{U}(0,1)$ &
$0.208^{+0.136}_{-0.137}$ \\

$\rho_{\star}$ (g\,cm$^{-3}$) &
$\mathcal{G}(1.46,0.10)$ &
$1.491^{+0.093}_{-0.103}$ &
$\mathcal{G}(1.72,0.17)$ &
$1.718^{+0.134}_{-0.140}$ \\

\hline
\multicolumn{5}{l}{Planet c $-$ sampled parameters} \\
\hline
$T_{0,\mathrm{c}}$ (RJD) &
$\mathcal{U}(9750,10150)$ &
$9798.2^{+5.6}_{-6.3}$ &
$\mathcal{U}(8933.70,8933.76)$ &
$8933.7337^{+0.0055}_{-0.0045}$ \\

$P_{\mathrm{c}}$ (d) &
$\mathcal{U}(2,500)$ &
$235.8^{+9.2}_{-8.5}$ &
$\mathcal{U}(4.023,4.025)$ &
$4.0239478^{+0.0000036}_{-0.0000034}$ \\

$R_{\mathrm{p,c}}/R_{\star}$ &
... & ... &
$\mathcal{U}(0,0.1)$ &
$0.02570^{+0.00107}_{-0.00104}$ \\

$K_{\mathrm{c}}$ (m\,s$^{-1}$) &
$\mathcal{U}(0,100)$ &
$45.9 \pm 1.4$ &
$\mathcal{U}(0,100)$ &
$2.67 \pm 0.50$ \\

$b_{\mathrm{c}}$ &
... & ... &
$\mathcal{U}(0,1)$ &
$0.312^{+0.138}_{-0.205}$ \\

\hline
\multicolumn{5}{l}{Limb-darkening parameters} \\
\hline
$q_1$ &
$\mathcal{G}(0.3240,0.0145)$ &
$0.3249 \pm 0.0145$ &
$\mathcal{G}(0.3599,0.0160)$ &
$0.3618 \pm 0.0158$ \\

$q_2$ &
$\mathcal{G}(0.3596,0.0070)$ &
$0.3598 \pm 0.0070$ &
$\mathcal{G}(0.3872,0.00775)$ &
$0.3876 \pm 0.0074$ \\

\hline
\multicolumn{5}{l}{Offsets and jitter terms} \\
\hline
$\gamma_{\mathrm{RV}}$ (m\,s$^{-1}$) &
$\mathcal{U}(-14175.4,-13819.3)$ &
$-14007.8^{+4.8}_{-5.0}$ &
$\mathcal{U}(-8500,-8300)$ &
$-8425.0 \pm 7.6$ \\

$\gamma_{\mathrm{H\alpha}}$ &
$\mathcal{U}(0.1915,0.4217)$ &
$0.3063^{+0.0026}_{-0.0027}$ &
$\mathcal{U}(0.2,0.4)$ &
$0.2894 \pm 0.0029$ \\

$\gamma_{\mathrm{BIS}}$ (m\,s$^{-1}$) &
$\mathcal{U}(-112.5,157.7)$ &
$13.87^{+0.88}_{-0.90}$ &
... & ... \\

$\sigma_{\mathrm{RV}}$ (m\,s$^{-1}$) &
$\mathcal{M}(0.5,10)$ &
$2.25^{+0.88}_{-0.69}$ &
$\mathcal{M}(0.5,10)$ &
$1.49^{+0.50}_{-0.49}$ \\

$\sigma_{\mathrm{H\alpha}}$ &
$\mathcal{M}(0.0005,0.0100)$ &
$0.00222^{+0.00033}_{-0.00030}$ &
$\mathcal{M}(0.0005,0.01)$ &
$0.00201 \pm 0.00043$ \\

$\sigma_{\mathrm{BIS}}$ (m\,s$^{-1}$) &
$\mathcal{M}(0.5,10)$ &
$5.10^{+0.69}_{-0.62}$ &
... & ... \\

$\sigma_{\mathrm{LC}}$ &
$\mathcal{U}(0.0000,0.0100)$ &
$0.000308^{+0.000013}_{-0.000014}$ &
$\mathcal{U}(0,100)$ &
$0.000205^{+0.000024}_{-0.000017}$ \\

\hline
\multicolumn{5}{l}{Gaussian process hyperparameters (quasi-periodic)} \\
\hline
$A_1$ (m\,s$^{-1}$) &
$\mathcal{U}(-100,100)$ &
$13.6^{+3.5}_{-2.4}$ &
$\mathcal{U}(0,100)$ &
$19.7^{+6.0}_{-5.2}$ \\

$B_1$ (m\,s$^{-1}$\,d$^{-1}$) &
$\mathcal{U}(0,100)$ &
$35.2^{+8.4}_{-5.8}$ &
... & ... \\

$A_2$ &
$\mathcal{U}(0,0.5)$ &
$0.0076^{+0.0018}_{-0.0013}$ &
$\mathcal{U}(0,0.5)$ &
$0.00729 \pm 0.00179$ \\

$A_3$ (m\,s$^{-1}$) &
$\mathcal{U}(-100,100)$ &
$0.75^{+1.14}_{-1.10}$ &
... & ... \\

$B_3$ (m\,s$^{-1}$\,d$^{-1}$) &
$\mathcal{U}(-100,100)$ &
$-28.6^{+4.8}_{-6.9}$ &
... & ... \\

$\lambda_{\mathrm{e}}$ (d) &
$\mathcal{U}(10,100)$ &
$20.8^{+5.1}_{-3.7}$ &
$\mathcal{U}(10,500)$ &
$65.6 \pm 24.0$ \\

$\lambda_{\mathrm{p}}$ &
$\mathcal{U}(0.1,2.0)$ &
$0.477^{+0.050}_{-0.042}$ &
$\mathcal{U}(0.1,10)$ &
$1.18 \pm 0.21$ \\

$P_{\mathrm{GP}}$ (d) &
$\mathcal{U}(10.0,13.5)$ &
$12.37 \pm 0.10$ &
$\mathcal{U}(20,30)$ &
$24.00^{+0.47}_{-0.46}$ \\

\hline
\multicolumn{5}{l}{Planet b $-$ derived parameters} \\
\hline
$R_{\mathrm{p,b}}$ ($R_{\oplus}$) & (derived) & $2.19 \pm 0.09$ & (derived) & $2.27 \pm 0.10$ \\
$M_{\mathrm{p,b}}$ ($M_{\oplus}$) & (derived) & $6.7 \pm 1.8$ & (derived) & $7.10 \pm 0.78$ \\
$\rho_{\mathrm{p,b}}$ (g\,cm$^{-3}$) & (derived) & $3.52^{+1.10}_{-1.00}$ & (derived) & $3.34^{+0.64}_{-0.54}$ \\
$g_{\mathrm{p,b}}$ ($10^{3}$\,cm\,s$^{-2}$) & (derived) & $1.37^{+0.40}_{-0.38}$ & (derived) & $1.35^{+0.20}_{-0.18}$ \\
$a_{\mathrm{b}}/R_{\star}$ & (derived) & $5.16^{+0.11}_{-0.12}$ & (derived) & $5.69 \pm 0.15$ \\
$a_{\mathrm{b}}$ (au) & (derived) & $0.02374 \pm 0.00088$ & (derived) & $0.02355 \pm 0.00009$ \\
$i_{\mathrm{b}}$ (deg) & (derived) & $87.0 \pm 1.2$ & (derived) & $87.80^{+1.20}_{-1.32}$ \\
$T_{\mathrm{eq,b}}$ (K) & (derived) & $1790^{+28}_{-27}$ & (derived) & $1585 \pm 33$ \\
$F_{\mathrm{p,b}}$ ($F_{\oplus}$) & (derived) & $1711^{+109}_{-100}$ & (derived) & $1051^{+91}_{-86}$ \\
$\Delta_{\mathrm{b}}$ (ppt) & (derived) & $0.411 \pm 0.021$ & (derived) & $0.546^{+0.035}_{-0.034}$ \\
$T_{14,\mathrm{b}}$ (h) & (derived) & $1.93^{+0.07}_{-0.08}$ & (derived) & $1.91^{+0.07}_{-0.08}$ \\

\hline
\multicolumn{5}{l}{Planet c $-$ derived parameters} \\
\hline
$R_{\mathrm{p,c}}$ ($R_{\oplus}$) & ... & ... & (derived) & $2.50 \pm 0.13$ \\
$M_{\mathrm{p,c}}$ ($M_{\oplus}$) & ... & ... & (derived) & $6.0 \pm 1.1$ \\
$M_{\mathrm{p,c}}\sin i_{\mathrm{c}}$ ($M_{\oplus}$) & (derived) & $444^{+18}_{-17}$ & ... & ... \\
$\rho_{\mathrm{p,c}}$ (g\,cm$^{-3}$) & ... & ... & (derived) & $2.11^{+0.56}_{-0.47}$ \\
$g_{\mathrm{p,c}}$ ($10^{3}$\,cm\,s$^{-2}$) & ... & ... & (derived) & $0.94^{+0.21}_{-0.19}$ \\
$a_{\mathrm{c}}/R_{\star}$ & (derived) & $162.9^{+5.8}_{-5.4}$ & (derived) & $11.37 \pm 0.30$ \\
$a_{\mathrm{c}}$ (au) & (derived) & $0.747 \pm 0.019$ & (derived) & $0.04708 \pm 0.00018$ \\
$i_{\mathrm{c}}$ (deg) & ... & ... & (derived) & $88.39^{+0.80}_{-0.84}$ \\
$T_{\mathrm{eq,c}}$ (K) & (derived) & $319.2 \pm 7.2$ & (derived) & $1121 \pm 24$ \\
$F_{\mathrm{p,c}}$ ($F_{\oplus}$) & (derived) & $1.73 \pm 0.16$ & (derived) & $263^{+23}_{-21}$ \\
$\Delta_{\mathrm{c}}$ (ppt) & ... & ... & (derived) & $0.660^{+0.055}_{-0.053}$ \\
$T_{14,\mathrm{c}}$ (h) & ... & ... & (derived) & $2.63^{+0.13}_{-0.20}$ \\

\end{longtable}
\tablefoot{Posterior values correspond to the median and 68\% credible intervals. The adopted baseline models assume circular orbits for all planets. The reported ephemeris of TOI-426~c is conditional on the adopted circular parametrisation (Sect.~\ref{sec:TOI-426}). RJD $\equiv$ BJD$-2450000$. Priors: $\mathcal{U}(a,b)$ uniform, $\mathcal{G}(\mu,\sigma)$ Gaussian, and $\mathcal{M}(a,b)$ modified Jeffreys.}

\begin{table}[h]
\renewcommand{\arraystretch}{1.15}
\setlength{\tabcolsep}{9.2pt}
\parbox{.47\linewidth}{
\centering
\caption{TESS SAP photometry of TOI-426.}
\begin{tabular}{ccc}
\hline \hline
BJD$_{\rm TDB}$ (d) & SAP flux ($\mathrm{e^{-}\,s^{-1}}$) & Sector \\
\hline
2458437.993 & $23376 \pm 27$ & TESS05 \\
$\cdots$ & $\cdots$ & $\cdots$ \\
\hline
\end{tabular}
\label{tab:TESS_data_426}
\tablefoot{The full table is available at the CDS.}
}
\hfill
\parbox{.47\linewidth}{
\centering
\caption{TESS SAP photometry of TOI-1839.}
\begin{tabular}{ccc}
\hline \hline
BJD$_{\rm TDB}$ (d) & SAP flux ($\mathrm{e^{-}\,s^{-1}}$) & Sector \\
\hline
2458930.190 & $13409 \pm 18$ & TESS23 \\
$\cdots$ & $\cdots$ & $\cdots$ \\
\hline
\end{tabular}
\label{tab:TESS_data_1839}
\tablefoot{The full table is available at the CDS.}
}
\end{table}

\begin{table*}[h]
\renewcommand{\arraystretch}{1.15}
\centering
\scriptsize
\caption{HARPS radial velocities and activity indicators of TOI-426.}
\label{tab:HARPS_data_426}
\resizebox{\textwidth}{!}{
\begin{tabular}{lllllllll}
\hline \hline
BJD$_{\rm TDB}$ (d) & RV ($\mathrm{m\,s^{-1}}$) & FWHM ($\mathrm{m\,s^{-1}}$) & BIS ($\mathrm{m\,s^{-1}}$) & Contrast (\%) & $S$ index & H$\alpha$ & NaD & Ca \\
\hline
2459847.805 
& $-14037.7 \pm 2.5$ 
& $8274 \pm 12$ 
& $17.5 \pm 3.5$ 
& $43.130$ 
& $0.3286 \pm 0.0052$ 
& $0.3174 \pm 0.0022$ 
& $0.2815 \pm 0.0013$ 
& $0.2662 \pm 0.0026$ \\
$\cdots$ 
& $\cdots$ 
& $\cdots$ 
& $\cdots$ 
& $\cdots$ 
& $\cdots$ 
& $\cdots$ 
& $\cdots$ 
& $\cdots$ \\
\hline
\end{tabular}
}
\tablefoot{The full table is available at the CDS.}
\end{table*}

\begin{table*}[h]
\renewcommand{\arraystretch}{1.15}
\centering
\scriptsize
\caption{HARPS radial velocities and activity indicators of TOI-1839.}
\label{tab:HARPS_data_1839}
\resizebox{\textwidth}{!}{
\begin{tabular}{lllllllll}
\hline \hline
BJD$_{\rm TDB}$ (d) & RV ($\mathrm{m\,s^{-1}}$) & FWHM ($\mathrm{m\,s^{-1}}$) & BIS ($\mathrm{m\,s^{-1}}$) & Contrast (\%) & $S$ index & H$\alpha$ & NaD & Ca \\
\hline
2459223.871 
& $-8433.2 \pm 1.6$ 
& $7151 \pm 10$ 
& $-4.8 \pm 2.3$ 
& $53.486$ 
& $0.2477 \pm 0.0054$ 
& $0.2719 \pm 0.0017$ 
& $0.2119 \pm 0.0010$ 
& $0.2000 \pm 0.0025$ \\
$\cdots$ 
& $\cdots$ 
& $\cdots$ 
& $\cdots$ 
& $\cdots$ 
& $\cdots$ 
& $\cdots$ 
& $\cdots$ 
& $\cdots$ \\
\hline
\end{tabular}
}
\tablefoot{The full table is available at the CDS.}
\end{table*}

\begin{table}[h]
\renewcommand{\arraystretch}{1.15}
\setlength{\tabcolsep}{9.2pt}
\parbox{.47\linewidth}{
\centering
\caption{WASP photometry of TOI-426.}
\begin{tabular}{ccc}
\hline \hline
HJD$_{\rm UTC}$ (d) & Flux & Error \\
\hline
2454743.496 & $0.0208$ & $0.0042$ \\
$\cdots$ & $\cdots$ & $\cdots$ \\
\hline
\end{tabular}
\label{tab:WASP_data_426}
\tablefoot{The full table is available at the CDS.}
}
\hfill
\parbox{.47\linewidth}{
\centering
\caption{WASP photometry of TOI-1839.}
\begin{tabular}{ccc}
\hline \hline
HJD$_{\rm UTC}$ (d) & Flux & Error \\
\hline
2454857.776 & $-0.0037$ & $0.0084$ \\
$\cdots$ & $\cdots$ & $\cdots$ \\
\hline
\end{tabular}
\label{tab:WASP_data_1839}
\tablefoot{The full table is available at the CDS.}
}
\end{table}

\end{appendix}

\end{document}